%% file: ArXiV-2026-Radez-Bohak.tex
\documentclass{article}

\usepackage{arxiv}

\input{packages}
\input{commands}

\title{Empowering \acs{npc} Dialogue with Environmental Context Using \acsp{llm} and Panoramic Images}
\author{Grega Radež\\
	Faculty of Computer and Information Science,\\
    University of Ljubljana,\\
    Večna pot 113,\\
    1000 Ljubljana, Slovenia\\
	\texttt{gr1561@student.uni-lj.si} \\
	\And
	\href{https://orcid.org/0000-0002-9015-2897}{\includegraphics[scale=0.06]{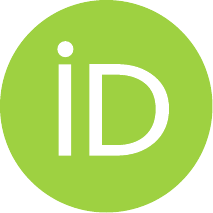}\hspace{1mm}Ciril Bohak} \\
	Faculty of Computer and Information Science,\\
    University of Ljubljana,\\
    Večna pot 113,\\
    1000 Ljubljana, Slovenia\\
	\texttt{ciril.bohak@fri.uni-lj.si} \\
}

\date{}

\renewcommand{\headeright}{Technical Report}
\renewcommand{\undertitle}{Technical Report}
\renewcommand{\shorttitle}{Empowering \acsp{npc} with Environmental Context Using \acsp{llm}}

\hypersetup{
pdftitle={Empowering NPC Dialogue with Environmental Context Using LLMs and Panoramic Images},
pdfsubject={},
pdfauthor={Grega Radež, Ciril Bohak},
pdfkeywords={Context-Aware NPC, Immersive Gameplay, Environmental Perception, Semantic Perception},
}

\begin{document}
\maketitle

\input{content/00-abstract}

\input{content/01-introduction}
\input{content/02-related-work}
\input{content/03-method}
\input{content/04-results}
\input{content/05-discussion}
\input{content/06-conclusion}

\balance
\bibliographystyle{unsrtnat}
\bibliography{references}

\newpage

\maketitle
\input{content/appendix}

\end{document}

%% file: packages.tex
\usepackage[utf8]{inputenc} 
\usepackage[english]{babel} 

\usepackage{tikz}
\usepackage{graphicx}
\usepackage[T1]{fontenc}    
\usepackage[unicode,hidelinks]{hyperref}       
\usepackage{url}            
\usepackage{amsfonts}       
\usepackage{amsmath} 
\usepackage{float}
\usepackage{enumitem}
\usepackage[ruled,vlined]{algorithm2e}
\usepackage{caption}
\usepackage{subcaption}
\usepackage{pgfplots}
\pgfplotsset{width=10cm,compat=1.9}
\usepgfplotslibrary{external}
\usepackage{booktabs}
\usepackage{multirow}
\usepackage{acro}
\usepackage{microtype}
\usepackage{nicefrac}       
\usepackage{natbib}
\usepackage{doi}
\usepackage{balance}

\usepackage[nameinlink,capitalize]{cleveref}
\crefformat{footnote}{#2\textsuperscript{\normalfont #1}#3}

\usepackage{placeins} 

%% file: commands.tex
\newcommand{\etal}{\emph{et al.}}
\newcommand{\eg}{\emph{e.g., }}

\DeclareAcronym{llm}{
    short = {LLM},
    long = {large language model}
}

\DeclareAcronym{npc}{
    short = {NPC},
    long = {non-playable character}
}

\DeclareAcronym{ai}{
    short = {AI},
    long = {artificial inteligence}
}

\DeclareAcronym{pcg}{
    short = {PCG},
    long = {procedural content generation}
}

\DeclareAcronym{fcn}{
    short = {FCN},
    long = {fully convolutional network}
}

\DeclareAcronym{rl}{
    short = {RL},
    long = {reinforcement learning}
}

\DeclareAcronym{ram}{
    short = {RAM++},
    long = {Recognize Anything Model}
}

\DeclareAcronym{ue}{
    short = {UE5},
    long = {Unreal Engine 5}
}

\DeclareAcronym{cv}{
    short = {CV},
    long = {Computer Vision}
}

\DeclareAcronym{sam}{
    short = {SAM},
    long = {Segment Anything Model}
}

%% file: content/00-abstract.tex
\begin{abstract}
We present an approach for enhancing non-playable characters (NPCs) in games by combining large language models (LLMs) with computer vision to provide contextual awareness of their surroundings. Conventional NPCs typically rely on pre-scripted dialogue and lack spatial understanding, which limits their responsiveness to player actions and reduces overall immersion. Our method addresses these limitations by capturing panoramic images of an NPC’s environment and applying semantic segmentation to identify objects and their spatial positions. The extracted information is used to generate a structured JSON representation of the environment, combining object locations derived from segmentation with additional scene graph data within the NPC’s bounding sphere, encoded as directional vectors. This representation is provided as input to the LLM, enabling NPCs to incorporate spatial knowledge into player interactions. As a result, NPCs can dynamically reference nearby objects, landmarks, and environmental features, leading to more believable and engaging gameplay. We describe the technical implementation of the system and evaluate it in two stages. First, an expert interview was conducted to gather feedback and identify areas for improvement. After integrating these refinements, a user study was performed, showing that participants preferred the context-aware NPCs over a non-context-aware baseline, confirming the effectiveness of the proposed approach.
\keywords{Context-Aware \acsp{npc} \and Immersive Gameplay \and Environmental Perception \and Semantic Perception}
\end{abstract}

%% file: content/01-introduction.tex
\section{Introduction}
\label{sec:introduction}
Well designed \acfp{npc} in video games play a central role in player engagement and narrative immersion. These characters make immersion deeper and the virtual worlds more believable by deepening the storyline and providing players with a sense of agency through interactive dialogue. However, despite remarkable advancements in game technology, game design, and \ac{ai} for games, \acs{npc} interactions frequently remain constrained by pre-scripted dialogue trees and predictable behavior patterns. Such static systems, while functional in specific contexts, fail to fully leverage the dynamic and richly detailed environments that modern game worlds now offer.

This disconnect between \acp{npc} and their surroundings often disrupts immersion. For instance, players may encounter \acp{npc} oblivious to environmental changes, such as weather, dynamic events, or even the consequences of the player's actions. These shortcomings can make the game world feel artificial and, in some cases, diminish the overall gaming experience~\cite{Isbister_2022}. While \ac{npc} \ac{ai} has undergone notable progress---particularly in terms of adaptive behavior and learning systems that respond to player actions~\cite{Laird_2001a,Laird_2001b}---most \acp{npc} still lack the ability to interpret and integrate real-time environmental context into their interactions. This limitation underscores a significant gap in the development of interactive and immersive game worlds.

The increasing complexity of game environments makes this challenge even harder. Modern games now feature open worlds teeming with dynamic elements such as real-time weather systems, detailed object physics, and non-linear storylines that unfold differently depending on player decisions. These advancements set the stage for heightened player expectations: \acp{npc} should no longer function merely as reactive agents tied to scripted scenarios but as proactive entities capable of perceiving, understanding, and responding to the environment in ways that enhance realism and narrative cohesion. For instance, an \ac{npc} that comments on a nearby event, such as the presence of a strange object in a village or the arrival of reinforcements during a battle, adds depth to the game world and makes the interaction feel meaningful.

This evolution toward contextually aware \acp{npc} aligns with broader trends in \ac{ai} and natural language processing. The emergence of \acfp{llm} offers new opportunities for transforming \ac{npc} dialogue systems. Unlike traditional rule-based methods, \acp{llm} excel at generating human-like responses and can be adapted to incorporate a wide range of contextual inputs, including semantic information about the game environment~\cite{cox2023conversational}. However, realizing the potential of these models requires novel approaches to bridge the gap between visual data and textual reasoning. Panoramic images, for example, can capture the richness and spatial dynamics of a \ac{npc}'s surroundings, providing an essential input for generating context-aware dialogue. By processing these images through semantic segmentation and object recognition techniques, it becomes possible to distill actionable insights about the environment into a format that \acp{llm} can interpret and use.

This paper argues that empowering \acp{npc} with environmental context is not merely an incremental improvement but a paradigm shift in game design. It enables a new class of \acp{npc} capable of engaging players on a deeper level by dynamically reflecting the world around them. Such capabilities have implications beyond games, offering insights into human-computer interaction, virtual training simulations, and the development of immersive storytelling experiences. By integrating \acp{llm} with panoramic imaging and advanced computer vision techniques, we present a framework that bridges the gap between static dialogue systems and the next generation of contextually aware \acp{npc}.

In this paper, we propose a novel system that enhances the contextual awareness of \acp{npc}, offering the following key contributions:
\begin{itemize}
    \item \textbf{Adding environmental awareness to \acp{npc} {thro\-ugh} panoramic image analysis:} We utilize semantic segmentation on panoramic images to extract environmental data, translating complex visual inputs into structured information that can be effectively processed and understood by \acs{llm}.
    
    \item \textbf{Development of context-sensitive \acp{npc}:} By integrating environmental context, our system enables \acp{npc} to reference nearby objects, landmarks, and spatial relationships during their interactions with players. This approach introduces dynamic, {envi\-ron\-ment}-aware conversations that surpass traditional static dialogue trees or basic \ac{llm}-based chatbot behaviors.

    \item \textbf{Enhancing realism in \ac{npc} behavior:} Our work bridges the gap between the realism of game worlds and the responsiveness of \ac{npc} interactions, empowering \acp{npc} to ``perceive'' and interpret their surroundings. This capability lays the groundwork for more immersive, contextually relevant dialogue systems in gaming and beyond.
\end{itemize}

In the remainder of the paper, we first present related work in \Cref{sec:related-work}, followed by the presentation of our method in \Cref{sec:method}. Next, we present the results in \Cref{sec:results} and discussion in \Cref{sec:discussion}. In \Cref{sec:conclusion}, we present the conclusions and the possible future work. 

%% file: content/02-related-work.tex
\section{Related Work}
\label{sec:related-work}
Enhancing \acp{npc} with more realistic, adaptive, and believable behavior has long been a central topic in game \ac{ai} and interactive narrative research. While significant progress has been made in adaptive behavior modeling, procedural dialogue, and narrative generation, relatively few approaches incorporate real-time perception of the game environment into \ac{npc} reasoning and interaction. This section reviews relevant research on \ac{npc} dialogue systems, adaptive game \ac{ai}, environmental perception, context-aware interaction, and recent uses of \acp{llm} and \acf{cv} in games, situating our work within these domains.

\subsection{\ac{npc} Dialogue Systems}

Traditional \ac{npc} dialogue systems rely predominantly on pre-defined branching dialogue trees, offering limited flexibility and variation based on player choices. Early work in computational narrative, such as the story understanding model by Black and Wilensky~\cite{Black_1980}, focused on dynamic story generation but remained purely narrative-driven, with no awareness of the physical game environment.

A major step toward more dynamic interaction was \emph{Façade}, developed by Mateas and Stern~\cite{Mateas_2003}, which enabled \acp{npc} to respond to player actions within an emergent narrative framework. Despite this advancement, the \acp{npc} remained bound to narrative logic and internal state, lacking any explicit perception or understanding of the surrounding environment. Feng~\etal~\cite{Feng2016} further explored autonomous behavior learning for \acp{npc}, yet environmental cues were not incorporated into the decision-making process.

Feng~\cite{Feng_2014} presents a comprehensive investigation into \ac{npc} behavior modeling through imitative learning, online adaptation, and affective modeling. His work examines how \acp{npc} can learn behavior patterns from players, adapt through interaction and feedback in dynamic settings, and integrate emotion to create more believable characters. While this research substantially advances autonomy and affective behavior, the proposed models operate primarily on internal state and interaction feedback, without explicitly incorporating perception or structured representations of the surrounding game environment.

Basart~\cite{Basart_2022} proposes a narrative-centric approach emphasizing thematic coherence and responsiveness over traditional Aristotelian narrative structures. Although this approach dynamically adapts narratives to player input, environmental perception remains outside its scope.

\subsection{Adaptive \acp{npc} and Game \ac{ai}}

Adaptive game \ac{ai} has investigated methods enabling \acp{npc} to adjust behavior based on player actions and internal states. Laird’s anticipatory \acp{npc}~\cite{Laird_2001b} demonstrated how characters could predict player behavior and adapt accordingly, marking an important milestone in adaptive behavior design. However, adaptation was driven by internal models rather than perception of the physical game world.

Yannakakis and Togelius~\cite{Yannakakis_2011} explored experience-driven \ac{pcg}, allowing game environments and \ac{npc} behavior to co-evolve based on player interaction. Similarly, search-based \ac{pcg} approaches~\cite{Togelius_2010} generate environments tailored to player preferences. While these methods enhance adaptivity and variability, they do not enable \acp{npc} to explicitly perceive or reason about their surroundings during interaction.

More recently, Csepregi~\cite{Csepregi_2021} demonstrated that context-aware \ac{llm}-based \acp{npc} can improve player engagement through more dynamic dialogue. Vidrih and Mayahi~\cite{Vidrih_2023} highlight similar benefits of generative AI in creating personalized experiences outside gaming. These works underline the potential of \acp{llm} but primarily focus on dialogue generation rather than grounding interactions in real-time environmental perception.

\subsection{Environmental Awareness in Game \ac{ai}}

Environmental perception has been extensively studied in robotics and autonomous systems. Thrun~\etal~\cite{Thrun_2006} demonstrated how real-time sensor data enables autonomous navigation in complex environments. Despite their success, such perception techniques have been only sparsely applied to \acp{npc} in games, particularly for conversational interaction.

Advances in \ac{cv} have enabled fine-grained scene understanding. Long~\etal~\cite{Long_2015} introduced fully convolutional networks for semantic segmentation, allowing pixel-level scene interpretation. Kirillov~\etal~\cite{Kirillov_2019} further introduced panoptic segmentation, combining semantic and instance segmentation to capture object identities and spatial layout. These developments provide a foundation for enabling \acp{npc} to visually perceive their surroundings.

Work on spatial reasoning in navigation, such as Marza~\etal~\cite{Marza_2022}, shows that reinforcement-learning agents equipped with auxiliary perception tasks can develop strong spatial understanding without explicit maps. This emphasis on perception and spatial reasoning aligns closely with our goal of enabling \acp{npc} to interpret and reference their environment during interaction.

\subsection{Context-Aware \ac{npc} Interaction}

Context-aware computing was introduced by Schilit~\etal~\cite{Schilit_1994} as a paradigm in which systems adapt behavior based on situational context, such as location or nearby objects. While widely adopted in mobile and ubiquitous computing, its application to \acp{npc} interaction remains limited.

Most game \ac{npc} systems still lack mechanisms for grounding dialogue in explicit environmental context. Our work addresses this gap by integrating panoramic vision and semantic segmentation to provide \acp{npc} with structured, real-time representations of their surroundings, enabling spatially grounded interaction.

\subsection{Use of \acp{llm} in Games and Game Development}

The rapid adoption of \acp{llm} has significantly influenced game development and game \ac{ai}. Gallotta~\etal~\cite{Gallotta2024} present a comprehensive survey and research roadmap for \acp{llm} in games, covering narrative generation, player modeling, mixed-initiative design, and tooling. Sweetser~\cite{Sweetser2024} provides a preliminary scoping review highlighting emerging design patterns and open challenges related to agency, authorship, and ethics. Yang~\etal~\cite{Yang2025} offer an updated scoping review focused on GPT-based models between 2020 and 2024, identifying strong emphasis on dialogue and prototyping, but limited exploration of grounding \acp{llm} in real-time game-world perception.

Beyond surveys, individual applications span game generation~\cite{Hu_2024}, procedural narrative~\cite{Buongiorno_2024}, multiplayer role-playing dialogue~\cite{Koomen_2023}, emergent storytelling~\cite{Peng_2024}, user perception of generated dialogue~\cite{Akoury_2023}, and collaborative gameplay with \ac{llm}-driven \acp{npc}~\cite{Rao_2024}. While these approaches integrate \acp{llm} into gameplay or development pipelines, they largely treat the game world symbolically or implicitly, without explicit environmental grounding.

\subsection{Use of \acl{cv} Approaches in Games}

Game engines have been used both as platforms for \ac{cv} research, such as UnrealCV~\cite{Qiu_2016}, and as sources of synthetic training data~\cite{Krahenbuhl_2018}. Conversely, \ac{cv} techniques have been employed to augment gameplay, including visual perception systems for \acp{npc}~\cite{Mahmoud_2014}, behavior analysis~\cite{Fink_2007}, and automated control or cheating~\cite{Nhu_2023}. Erdelyi~\cite{Erdelyi_2019} explored the use of \ac{cv} for playing existing games. Despite this diversity of applications, the use of \ac{cv} for improving \acp{npc} environmental awareness in conversational contexts remains largely unexplored.

\subsection{Image Segmentation}
\label{sec:segmentacija}

A key enabling technology for our work is the Segment Anything Model (SAM)~\cite{kirillov2023segany}, developed by Meta. SAM enables zero-shot segmentation of unseen objects and is supported by one of the largest segmentation datasets to date. Its flexibility has led to numerous derivative tools~\cite{modelzoo}.

We evaluated several SAM-based approaches, including Semantic Segment Anything~\cite{chen2023semantic} and the Label Anything Pipeline~\cite{yu2023interactive}. Ultimately, we selected the Recognize Anything Model (RAM)~\cite{zhang2023recognize}, which supports open-set recognition with high accuracy and efficient inference. RAM builds upon earlier SAM-derived systems such as Grounded SAM~\cite{ren2024grounded}, Tag2Text~\cite{huang2023tag2text}, and BLIP~\cite{li2022blip}. Its ability to generate reliable object annotations makes it well suited for extracting structured environmental context to support \acp{npc} interaction.

%% file: content/03-method.tex
\section{Method}
\label{sec:method}
\begin{figure}[t]
    \centering
    \includegraphics[width=\linewidth]{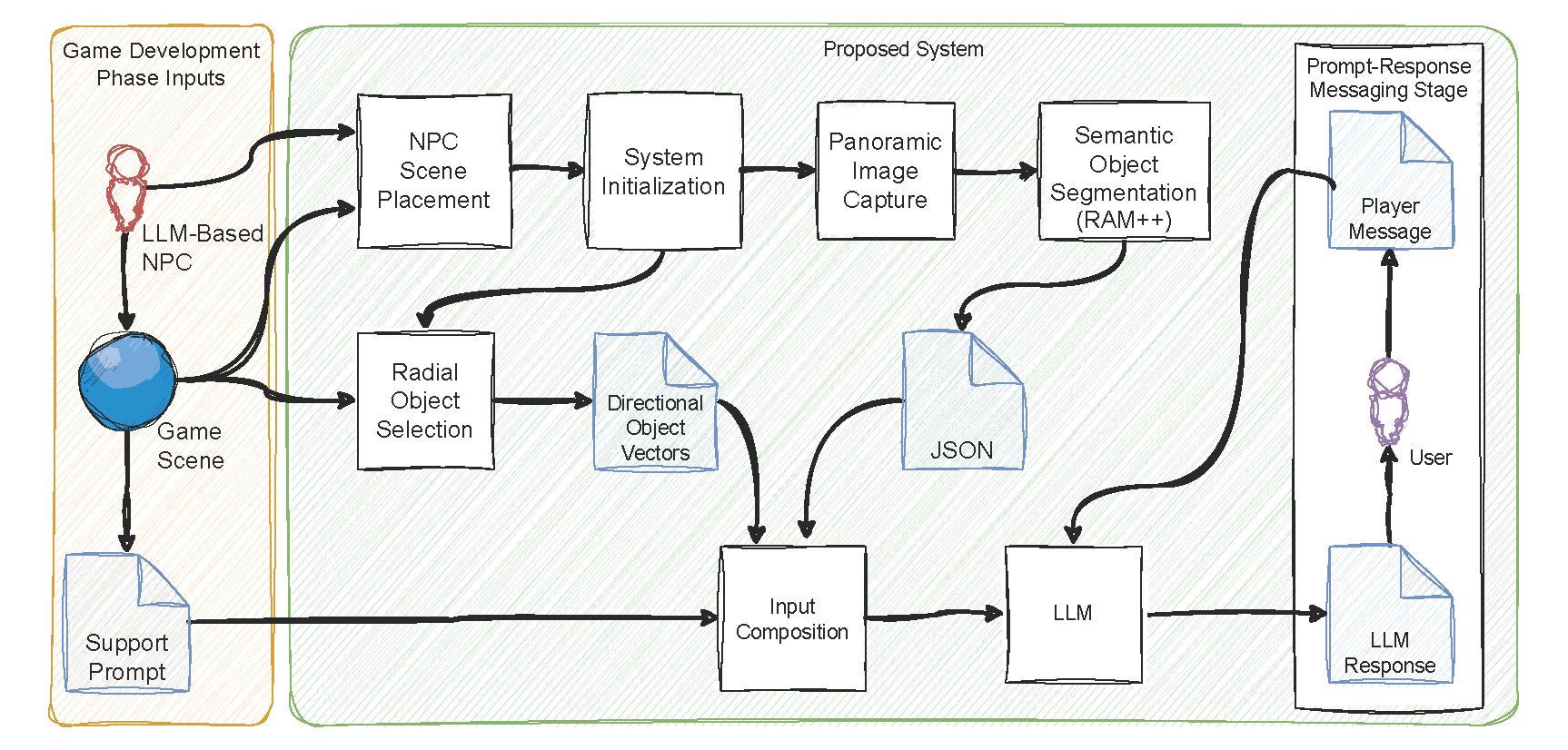}
    \caption{The proposed system structure. The yellow box present the inputs and outputs of the system, and the blocks in the green box present individual steps of our approach.}
    \label{fig:structure}
\end{figure}
The proposed approach was developed within the \ac{ue}\footnote{\url{https://www.unrealengine.com/}} developed by Epic Games Inc.\ with custom-built Blueprints and methods and is built on top of a \ac{llm}-based \ac{npc}, which has the context of its story defined either during training or with a \emph{Supporting Prompt} defining its context. Such \ac{npc} can be placed anywhere in the game \emph{Scene} but does not have any information on where it is and what is within its vicinity. During the game-building process, our plugin is initiated to provide the \ac{npc} with information about its surroundings using the data obtained with semantic segmentation of panoramic image captured at its location, a list of objects inside a bounding sphere with their directions, and a supporting prompt guiding the \ac{llm} to stay focused on the given job. The outline of our approach is presented in \Cref{fig:structure}. In the following subsections, we present details of each system component and their mutual connections.

\subsection{Inputs}
\label{sec:inputs}
The main input is a \ac{llm}-based \ac{npc}. In our case, this was a link to ChatGPT's\footnote{\url{https://chatgpt.com}} GPT-4 API by OpenAI\footnote{\url{https://openai.com}}. The \ac{llm} was fed a \emph{Supporting Prompt} presenting the \ac{npc}'s story. An example of such a supporting prompt is presented below.

\vspace{0.5cm}
\hrule
\vspace{0.1cm}
\noindent

\begin{sloppypar}
{\footnotesize \ttfamily
\noindent
You are a quest giver in a medieval fantasy world, {addre\-ssing} an {adven\-tu\-rer} who is standing directly in front of you and facing you. The {en\-vi\-ron\-ment} around\\ you contains various objects, and the adventurer seeks your wisdom in describing them.
You will be given objects with {di\-rec\-tio\-nal} vectors relative to your position. Your task is to convert these NPC-relative vectors into appropriate directions solely from the adventurer's perspective. Use the following guidelines:\\
- If the vector points to your left, describe the object as being "to your right."\\
- If the vector points to your right, describe the object as being "to your left."\\
- If the vector points in front of you, describe it as being "behind you."\\
- If the vector points behind you, describe it as being "in front of you."\\
- For height (z-axis), if the vector points upward, describe the object as "above," and if downward, describe it as "below."\\
Your descriptions should be rich, detailed, and atmospheric, fitting the tone of a medieval fantasy world. Avoid any modern or technical terms that break the immersion. Speak directly to the adventurer (the player character), and never address the player. Your tone should remain formal and reflective of a wise, ancient figure guiding an adventurer through a fantastical realm.
When given two sets of information—such as one in JSON format and another in plain text with directional vectors—some details may overlap or describe the same object in different ways. In these cases, use your best judgment to estimate what is most plausible or true, combining the information naturally into a cohesive description. If items from both sets are similar or refer to the same object, choose one description that best fits the context.\\
**Ensure your descriptions only refer to the adventurer's perspective. Do not include references to your own orientation in the description. When describing the direction of objects, feel free to mix references relevant to the adventurer's perspective, but avoid mentioning your own position.**
}
\end{sloppypar}
\vspace{0.1cm}
\hrule
\vspace{0.5cm}

\begin{figure*}[tb]
    \centering
    \includegraphics[width=\textwidth]{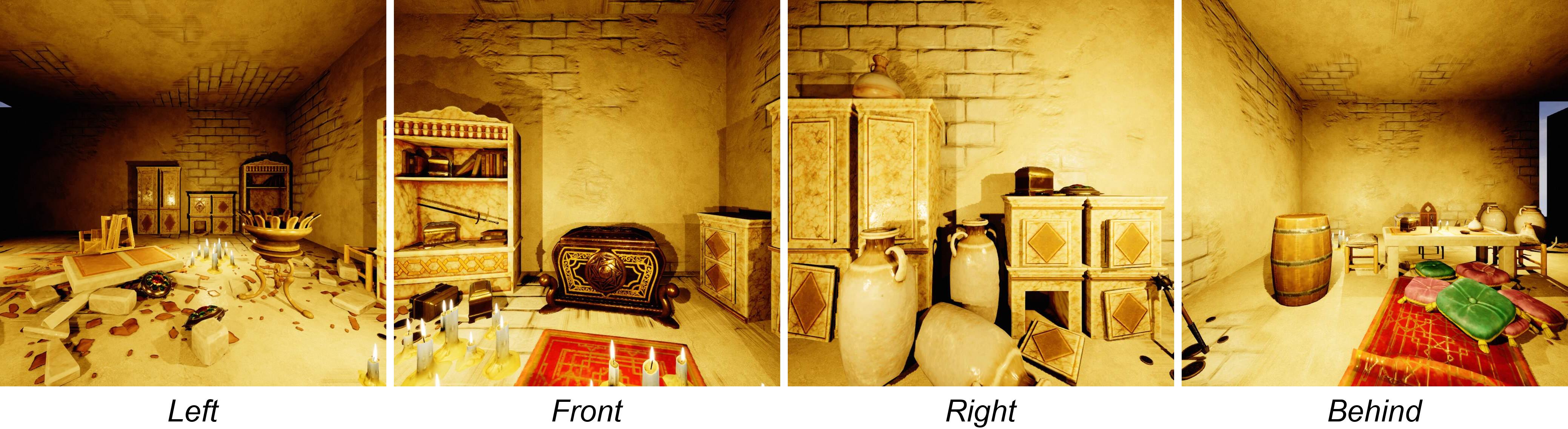}
    \caption{Panoramic image of an indoor scene composed of four images covering $90^{\circ}$ field of view, resulting in full $360^{\circ}$ panoramic image.}
    \label{fig:panorama}
\end{figure*}

The next input is the \emph{Scene}, assembled in \ac{ue}. An example of such an indoor scene is presented in~\Cref{fig:panorama}.

\subsection{System Initialization}
Our system is initialized during the game-building phase when the neighborhood contextual information is created. During the initialization step, the system activates the necessary plugins and builds the scene.

\subsection{Panoramic Image Capture}
In order to generate information used as inputs for the \ac{llm}-based \ac{npc}, we first need to gather data from the scene where the \ac{npc} is located. The first step is image capturing. We put a camera at eye-level of the \ac{npc} which takes a panoramic image composed of four images, each covering 90$^\circ$ of its view (\textit{front}, \textit{left}, \textit{right}, and \textit{behind}). We make sure the capturing process ignores objects and characters it does not or should not see (\eg portions of \ac{npc}'s own body). The obtained images are then fed as inputs for the next step: semantic object segmentation. An example of such an image set is shown in~\Cref{fig:panorama}.

\subsection{Semantic Object Segmentation using RAM++}
The images from the previous step are fed into the modified \ac{ram}~\cite{zhang2023recognize}. \ac{ram} is an open-set image tagging model effectively leveraging multi-grained text supervision and is used in the key stage of our method. We use four separate images instead of one continuous panoramic image to improve model performance since the model was trained with limited field-of-view images and not panoramic ones. We exploited this to include simple directions based on camera direction for the generated object tags. Generated tags were formatted in a JSON file format, with the tags being grouped into arrays \textit{left}, \textit{right}, \textit{front}, and \textit{behind}. The model is executed through a single Python script from within the \ac{ue}. It needs a pre-trained dataset accessible locally. An example of the resulting JSON file is presented below.

\vspace{0.5cm}
\hrule
\vspace{0.1cm}
\begin{sloppypar}
{\footnotesize \ttfamily
\noindent
\{\\
\indent "left": ["cabinet", "pottery", "closet", \ldots],\\
\indent "in-front": ["barrel", "basement", \ldots],\\
\indent "right": ["altar", "basement", "candle", \ldots],\\
\indent "behind": ["altar", "candle", \ldots]\\
\}
}
\end{sloppypar}
\vspace{0.1cm}
\hrule

\subsection{Radial Object Selection}
\label{sec:radial-object-selection}
Concurrently to \emph{System initialization} step, the system provides another flavor of context information from within the engine. We achieve this by using a component bounding sphere, which returns any object the sphere collides with within a given radius. The radius was empirically set to 10 meters. This stage is usable only when proper asset naming regulations exist within the game development environment. If the names of objects are meaningless, it is better to omit this step in the method. However, when objects are given proper names (or tags), this method can prove helpful, as it provides more precise directional data for each object and a better sense for plurality in the case of more than one instance of the same object. The directional vectors point to the object from the position of the \ac{npc}. An example of some of the data generated by this method is presented below.
\vspace{0.5cm}
\hrule
\vspace{0.1cm}
\noindent
\texttt{\small
Simple\char`_Shelf2, VEC:X=-0.940 Y=-0.340 Z=0.000\\
Simple\char`_Pot\char`_Stubby2, VEC:X=-0.456 Y=0.874 Z=-0.171\\
Barrel1, VEC:X=0.348 Y=0.937 Z=0.000
}
\vspace{0.1cm}
\hrule

\subsection{Prompt-Response Messaging Stage}
\begin{figure*}[t]
    \centering
    \includegraphics[width=\textwidth]{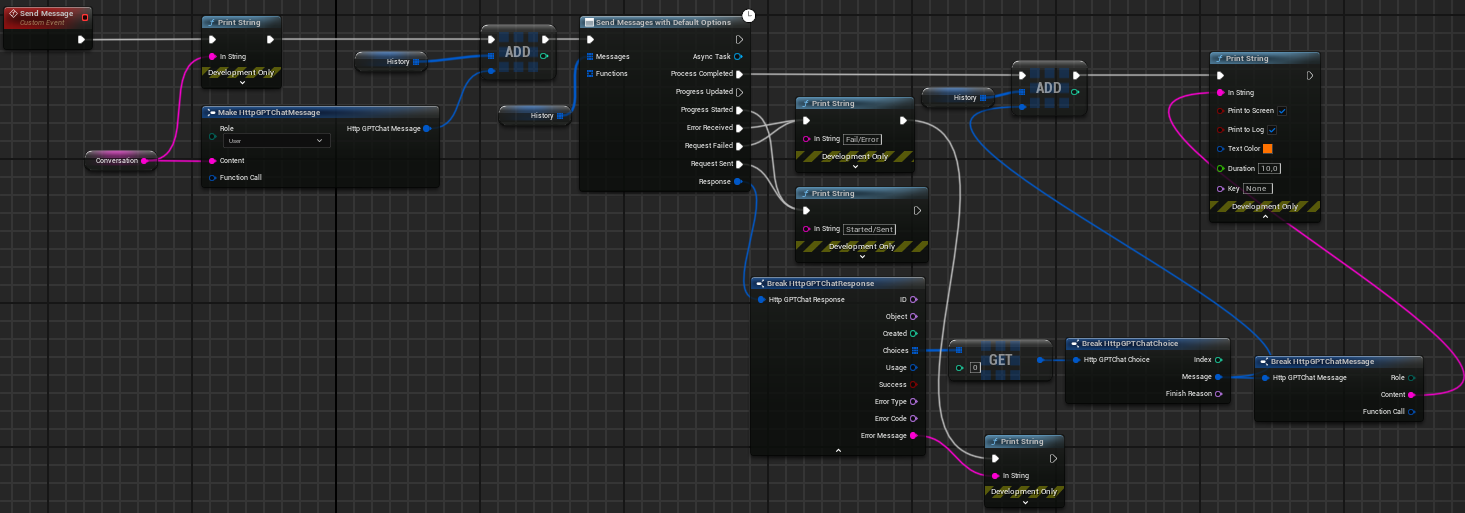}
    \caption{\ac{ue} blueprint layout of the Prompt-Response Messaging Stage.}
    \label{fig:prompt-response}
\end{figure*}

This stage lets us send messages to the chosen \ac{llm} and receive responses. It also tracks history to access previous conversations in the same simulation instance. The history gets cleared at the end of each instance and is not saved. The concrete implementation in the \ac{ue} blueprint system is shown \Cref{fig:prompt-response}.

\subsection{Input Composition Stage}
\begin{figure*}[t]
    \centering
    \includegraphics[width=\textwidth]{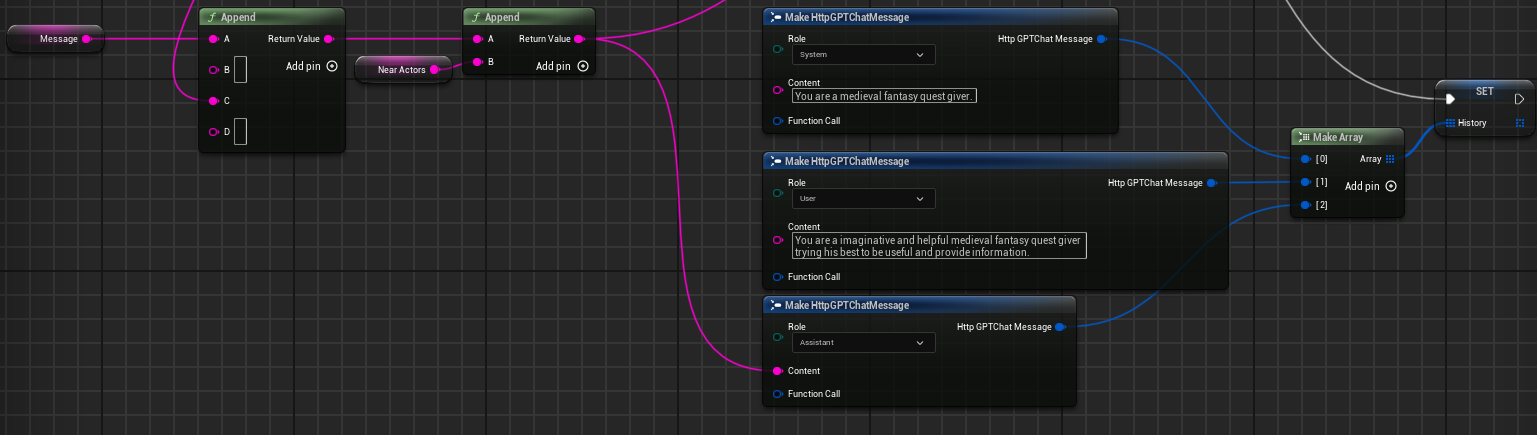}
    \caption{\ac{ue} blueprint layout of the \emph{Input Composition Stage}.}
    \label{fig:input-composition-stage}
\end{figure*}

This stage combines all the generated information regarding the world surrounding the \ac{npc} into a \emph{Supporting Prompt}. This prompt is then sent to \ac{llm} and saved as history. This is where the \ac{npc} gets the background information before the conversation starts. The \ac{ue} blueprint implementation is shown in \Cref{fig:input-composition-stage}.

\subsection{Player Interaction}
We try to provide a cohesive, immersive, and relatively detailed conversation experience. The player interacts with the system through conversational chat, similar to many web-based chatbot examples, by entering its prompt. The player's message is sent to the \ac{llm}, whose response is passed back to the player. Below is an example of such a conversation.

\vspace{0.5cm}
\hrule
\vspace{0.1cm}
\begin{sloppypar}
{\footnotesize \ttfamily
\noindent
\textbf{Player:} Hi, I am John Smith, an adventurer. Care to tell me what some of the things around you are?\\
\textbf{NPC:} Well met, brave adventurer John Smith. There are indeed many intriguing objects surrounding us. Let me share my knowledge. To your right, you would find numerous relics of convenience and history. There is a cabinet holding the secrets of the past and a drawer capturing the essence of time. Among them is a dresser that reminds me of the owner's habits and a variety of pots, each holding its own tale of the meals served in this humble abode.
}
\end{sloppypar}
\vspace{0.1cm}
\hrule

%% file: content/04-results.tex
\section{Evaluation and results}
\label{sec:results}
To test the presented system, we conducted two evaluations: (1) an interview with an expert and (2) a comparative user study where we compared our system with a baseline \ac{llm}-based system. We present the results in the following subsections.

\subsection{Interview with the Expert}
For the interview with the expert, we prepared four different tests on one scene to see how different inputs to the system affect the \ac{llm}-based \ac{npc}'s responses about its surroundings in the scene and how they compare to the control data. We compared the following scenarios:
\begin{enumerate}
    \item Test 1 -- All input data
    \item Test 2 -- Semantic Object Segmentation data only
    \item Test 3 -- Assistant Prompt only
    \item Test 4 -- Assistant Prompt and Radial Object Selection data
\end{enumerate}

We used the same user queries presented below in all test scenarios.
\vspace{0.1cm}
\hrule
\vspace{0.1cm}
\begin{sloppypar}
{\footnotesize \ttfamily
\noindent
\textbf{Q1:} Hi, I am John Smith, an adventurer. Care to tell me what some of the things around you are?\\
\textbf{Q2:} I am interested in this altar. Is there anything more you could tell me about it and its surroundings?\\
\textbf{Q3:} Do you have any idea what the objects near the altar could have been used for in combination with it? If not, can you help me theorise?
}
\end{sloppypar}
\vspace{0.1cm}
\hrule
\vspace{0.1cm}
In the following subsections, we present partial responses to the first question because the full responses are too long. The full responses are available in Supplemental Materials.

\subsubsection{Test 1 -- All Data Inputs}
This is the control test, which illustrates the envisioned final use case for our system. It uses all inputs: the (1) \textit{Radial Object Selection}, (2) \textit{Semantic Object Segmentation}, and (3) \textit{Support Prompt}. Below is the \ac{npc}'s answer to \textbf{\texttt{Q1}}.
\vspace{0.1cm}
\hrule
\vspace{0.1cm}
\begin{sloppypar}
{\footnotesize \ttfamily
\noindent
\textbf{A1:} Well met, brave adventurer John Smith. There are indeed many intriguing objects surrounding us. Let me share my knowledge. To your right, you would find numerous relics of convenience and history. There is a cabinet holding the secrets of the past and a drawer capturing the essence of time \ldots
}
\end{sloppypar}
\vspace{0.1cm}
\hrule

\subsubsection{Test 2 -- Semantic Object Segmentation data only}
This test uses \textit{Semantic Object Segmentation} data only, voiding the \textit{Support Prompt} and \textit{Radial Object Selection} data. An example of the segmentation data is shown in \Cref{sec:radial-object-selection}. The \ac{npc}'s answer to \texttt{\textbf{Q1}} is shown below.
\vspace{0.1cm}
\hrule
\vspace{0.1cm}
\begin{sloppypar}
{\footnotesize \ttfamily
\noindent
Greetings, John Smith, delighted that you've found your way to my humble¸abode. Allow me to enlighten you about some of the items here in my surroundings. To my left, you will find a rather ancient cabinet, a behemoth of oak and iron, with some pieces of delicate pottery precariously balanced atop it \ldots
}
\end{sloppypar}
\vspace{0.1cm}
\hrule

\subsubsection{Test 3 -- Support Prompt only}
This test uses the \textit{Support Prompt} only. It is shown in \Cref{sec:inputs}. This test was mainly done to see how the gaps of missing information are filled. The answer to \texttt{\textbf{Q1}} is shown below.
\vspace{0.1cm}
\hrule
\vspace{0.1cm}
\begin{sloppypar}
{\footnotesize \ttfamily
\noindent
To your left, you'll find an ancient oak tree, its branches reaching stark against the sky, begging the heavens for a taste of the immortality that it has been denied here on the soil. Its roots dive deep beneath the surface, gnarled and tangled. It has witnessed innumerable sunrises and sunsets, acted as a silent guardian for many a weary traveler seeking shelter under its durable canopy \ldots
}
\end{sloppypar}
\vspace{0.1cm}
\hrule
\vspace{0.1cm}

\subsubsection{Test 4 -- Supporting Prompt and Radial Object Selection data}
This test uses a combination of the (1) \textit{Support Prompt} and (2) data gathered with the \textit{Radial Object Selection} method shown in \Cref{sec:radial-object-selection}. The \ac{npc}'s answer to \texttt{\textbf{Q1}} with this data is shown below.
\vspace{0.1cm}
\hrule
\vspace{0.1cm}
\begin{sloppypar}
{\footnotesize \ttfamily
\noindent
Greetings, John Smith, and welcome to my humble abode. As a quest giver, it is often the adventurer's curiosity that steers the journey. Let me shed light on your surroundings: To your right you'll find numerous shelves, laden with a variety of pots. The simplest of these come in different forms: stubby, conical, fat, tall; each one with a different story to tell. Please exercise caution while handling these – some could be more delicate than they seem \ldots
}
\end{sloppypar}
\vspace{0.1cm}
\hrule
\vspace{0.1cm}

After showing the questions and answers of all the tests to the expert, we conducted an unstructured interview to receive feedback on the tests. The expert told us that he is most satisfied with the output of the fully implemented system. However, he was a bit confused about how to interpret the results since the panoramic image is captured from the perspective of the \ac{npc}, while the responses are formed for the user, facing the \ac{npc}. He agreed that we should conduct a comparative user study evaluating the system on multiple scenes. Since a study that would include all 4 test scenarios for multiple scenes would be very long, he helped us select two tests that should be compared: \textit{Test 1} -- fully implemented system and \textit{Test 3} -- support prompt only. Based on the feedback from the expert, we designed a user study, presented in the following section, together with the results.

\subsection{Comparative User Study}
To gather the responses of the general public for quantitative evaluation, we designed a user study in the form of an online questionnaire using Google Forms.

\begin{table*}[t]
\centering
\resizebox{\textwidth}{!}{
\begin{tabular}{ccccccccc}
& \multicolumn{4}{c}{\textbf{Scene 1}} & \multicolumn{4}{c}{\textbf{Scene 2}}\\
\cmidrule(lr){2-5}\cmidrule(lr){6-9}
 & \textbf{\texttt{Q1} ($\mu_R\uparrow$)} & \textbf{\texttt{Q2} ($\mu_R\uparrow$)} & \textbf{\texttt{Q3} ($\mu_R\uparrow$)} & \textbf{\texttt{Q4} ($\mu_R\uparrow$)} & \textbf{\texttt{Q1} ($\mu_R\uparrow$)} & \textbf{\texttt{Q2} ($\mu_R\uparrow$)} & \textbf{\texttt{Q3} ($\mu_R\uparrow$)} & \textbf{\texttt{Q4} ($\mu_R\uparrow$)}\\
\midrule
\textbf{\texttt{A1}} & 47 (3.65) & 52 (3.82) & 49 (\textbf{3.65}) & 45 (3.46) & 49 (\textbf{3.85}) & 42 (3.73) & 45 (3.66) & 42 (\textbf{3.88})\\ 
\textbf{\texttt{A2}} & 11 (\textbf{3.81}) & 6 (\textbf{3.83}) & 9 (3.11) & 13 (\textbf{3.53}) & 9 (3.33) & 16 (\textbf{3.75}) & 13 (\textbf{4.00}) & 16 (3.37) \\ 
\end{tabular}
}
\vspace{0.2cm}
\caption{Distribution of answer selection for the individual question and individual scene, where \textbf{\texttt{Q$i$}} is $i$-th query, \textbf{\texttt{A$j$}} is $j$-th answer, and $\mu_R$ is average answer rating on a 5-point Likert scale.}
\label{tab:answers}
\end{table*}

\subsection{Study Design}
The study compared two tests (1 and 3) using an altered supporting prompt presented in section 2.1 of the Supplemental Materials and included four query questions passed to both versions of the system. The questions are presented below:
\vspace{0.1cm}
\hrule
\vspace{0.1cm}
\begin{sloppypar}
{\footnotesize \ttfamily
\noindent
\textbf{Q1:} Hi, I'm John Smith, can you quickly describe the area we are in?\\
\textbf{Q2:} What is the most interesting thing to see here?\\
\textbf{Q3:} Can you tell me more about it and how it's connected to other things surrounding it?\\
\textbf{Q4:} Is there anything that looks like it doesn't belong in the area?
}
\end{sloppypar}
\vspace{0.1cm}
\hrule
\vspace{0.1cm}
The questionnaire was composed of the following parts:
\begin{enumerate}
    \item Introductory section presenting our system and our study.
    \item Basic demographic questions on age, gender, education, and field of study.
    \item Precise instructions on tasks and what is expected of the user.
    \item Panoramic image presenting \textit{Scene 1} -- an indoor scene.
    \item Queries \textbf{\texttt{Q1 - Q4}} with accompanying pairs of responses from both versions of the system.
    \item Selection of the more appropriate answer and its ranking on a 5-point Likert scale.
    \item Voluntary open-ended feedback for each query.
    \item steps 4 -- 7 for \textit{Scene 2} -- an outdoor scene.
    \item Voluntary feedback on the study and contact details for disseminating the study results.
\end{enumerate}
We distributed the questionnaire in several game development, gaming, and student social network groups and collected the responses over three weeks. The full questionnaire is available as part of the Supplementary Material.

\begin{figure*}[tb]
    \centering
    \includegraphics[width=\textwidth]{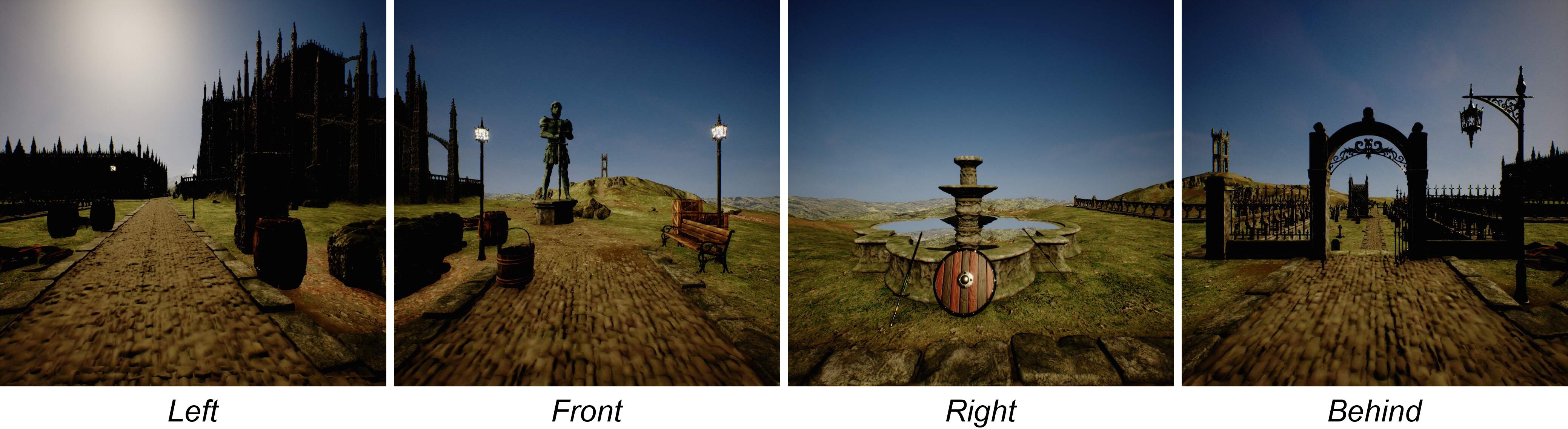}
    \caption{Panoramic image of an outdoor scene composed of four images covering $90^{\circ}$ field of view, resulting in full $360^{\circ}$ panoramic image.}
    \label{fig:panorama-outdoor}
\end{figure*}

We reused the indoor scene from the tests with the expert presented in \Cref{fig:panorama} and added an outdoor scene presented in \Cref{fig:panorama-outdoor}.

\subsubsection{Study Results}
Altogether, we gathered 58 responses.
\\

\noindent
\textbf{Demographics:} Participants consisted of 38 males and 15 females, and 5 participants did not want to share their gender. The participants were distributed into five age groups spanning 5 years as follows: 18--22 (2), 23--27 (25), 28--32 (6), 33--37 (1), and 38--42 (2). The education of participants was: High School (1), DiplEng (20), BSc (20), MSc (5), PhD (1), and Other (1). The distribution of their fields of study was Computer Science (41), Electrotechnics (3), Engineering (7), Arts (2), and Other (5). The bar charts showing the distributions are shown in \Cref{fig:graph3}.
\\

\newlength{\graphwidth}
\setlength{\graphwidth}{0.325\textwidth}

\newlength{\labeloffset}
\setlength{\labeloffset}{0.7cm}

\pgfplotsset{compat=1.18}

\begin{figure*}
     \centering
     \begin{subfigure}[b]{\graphwidth}
         \centering
         \resizebox{\graphwidth}{!}{
             \begin{tikzpicture}
                \begin{axis}[
                    title={},
                    title style={yshift=-7pt},
                    xbar,
                    xmax=55,
                    ytick=data,
                    enlarge y limits=0.2,
                    enlarge x limits=0.02,
                    bar width = 15pt,
                    ylabel={Years},
                    xlabel={\% of participants (\# of participants)},
                    xlabel style={yshift=-7pt},
                    symbolic y coords = {18-22, 23-27, 28-32, 33-37, 38-42},
                    font=\footnotesize,
                    nodes near coords,
                    nodes near coords style={
                        font=\footnotesize,
                        anchor=west,
                        /pgf/number format/fixed,
                        /pgf/number format/precision=2,
                        /pgf/number format/fixed zerofill
                    },
                ]
                    \addplot[
                        xbar,
                        fill=gray
                    ] coordinates {
                        (41.37,18-22) 
                        (43.10,23-27)
                        (10.34,28-32)
                        (1.72,33-37)
                        (3.44,38-42)
                    };
            
                    \node[anchor=west, font=\footnotesize] at (41.37,18-22) {\hspace{\labeloffset} (24)};
                    \node[anchor=west, font=\footnotesize] at (43.10,23-27) {\hspace{\labeloffset} (25)};
                    \node[anchor=west, font=\footnotesize] at (10.34,28-32) {\hspace{\labeloffset} (6)};
                    \node[anchor=west, font=\footnotesize] at (1.72,33-37) {\hspace{\labeloffset} (1)};
                    \node[anchor=west, font=\footnotesize] at (3.44,38-42) {\hspace{\labeloffset} (2)};
                \end{axis}
            \end{tikzpicture}
         }
         \caption{Age groups.}
         \label{fig:graph1}
     \end{subfigure}
     \hfill
     \begin{subfigure}[b]{\graphwidth}
         \centering
         \resizebox{\graphwidth}{!}{
            \begin{tikzpicture}
                \begin{axis}[title={},
                    title style={yshift=-7pt},
                    xbar,
                    xmax=65,
                    ytick=data,
                    enlarge y limits  = 0.2,
                    enlarge x limits  = 0.02,
                    bar width = 15pt,
                    ylabel={Education Level},
                    xlabel={\% of participants (\# of participants)},
                    xlabel style={yshift=-7pt},
                    symbolic y coords = {BSc, DipEng, MSc, HS, PhD, Other},
                    font=\footnotesize,
                    nodes near coords,
                    nodes near coords style={
                        font=\footnotesize,
                        anchor=west,
                        /pgf/number format/fixed,
                        /pgf/number format/precision=2,
                        /pgf/number format/fixed zerofill
                    },
                ]
                    \addplot[
                        xbar,
                        fill=gray
                    ] coordinates {
                                    (51.7,BSc)
                                    (34.5,DipEng)
                                    (8.6,MSc)
                                    (1.7,HS)
                                    (1.7,PhD)
                                    (1.7,Other)
                                    };
                
                    \node[anchor=west, font=\footnotesize] at (axis cs:51.72,BSc) {\hspace{\labeloffset} (30)};
                    \node[anchor=west, font=\footnotesize] at (axis cs:34.48,DipEng) {\hspace{\labeloffset} (20)};
                    \node[anchor=west, font=\footnotesize] at (axis cs:8.62,MSc) {\hspace{\labeloffset} (5)};
                    \node[anchor=west, font=\footnotesize] at (axis cs:1.72,HS) {\hspace{\labeloffset} (1)};
                    \node[anchor=west, font=\footnotesize] at (axis cs:1.72,PhD) {\hspace{\labeloffset} (1)};
                    \node[anchor=west, font=\footnotesize] at (axis cs:1.72,Other) {\hspace{\labeloffset} (1)};
                \end{axis}
            \end{tikzpicture}
         }
         \caption{Education Level.}
         \label{fig:graph2}
     \end{subfigure}
     \hfill
     \begin{subfigure}[b]{\graphwidth}
         \centering
         \resizebox{\graphwidth}{!}{
            \begin{tikzpicture}
                \begin{axis}[
                    title={},
                    title style={yshift=-7pt},
                    xbar,
                    xmax=90,
                    ytick=data,
                    enlarge y limits=0.2,
                    enlarge x limits=0.02,
                    bar width = 15pt,
                    ylabel={Field of Study},
                    xlabel={\% of participants (\# of participants)},
                    xlabel style={yshift=-7pt},
                    symbolic y coords = {CS, Electro., Eng., Arts, Other}, 
                    font=\footnotesize,
                    nodes near coords,
                    nodes near coords style={
                        font=\footnotesize,
                        anchor=west,
                        /pgf/number format/fixed,
                        /pgf/number format/precision=2,
                        /pgf/number format/fixed zerofill
                    },
                ]
                    
                    \addplot[
                        xbar,
                        fill=gray
                    ] coordinates {
                        (70.69,CS)
                        (5.17,Electro.)
                        (12.06,Eng.)
                        (3.44,Arts)
                        (8.62,Other)
                    };

                    \node[anchor=west, font=\footnotesize] at (axis cs:70.69,CS) {\hspace{\labeloffset} (41)};
                    \node[anchor=west, font=\footnotesize] at (axis cs:5.17,Electro.) {\hspace{\labeloffset} (3)};
                    \node[anchor=west, font=\footnotesize] at (axis cs:12.06,Eng.) {\hspace{\labeloffset} (7)};
                    \node[anchor=west, font=\footnotesize] at (axis cs:3.44,Arts) {\hspace{\labeloffset} (2)};
                    \node[anchor=west, font=\footnotesize] at (axis cs:8.62,Other) {\hspace{\labeloffset} (5)};
                \end{axis}
            \end{tikzpicture}
         }
         \caption{Field of Study.}
         \label{fig:graph3}
     \end{subfigure}
        \caption{Distribution of user study participants based on age (a), education level (b), and field of study (c).}
        \label{fig:graphs}
\end{figure*}
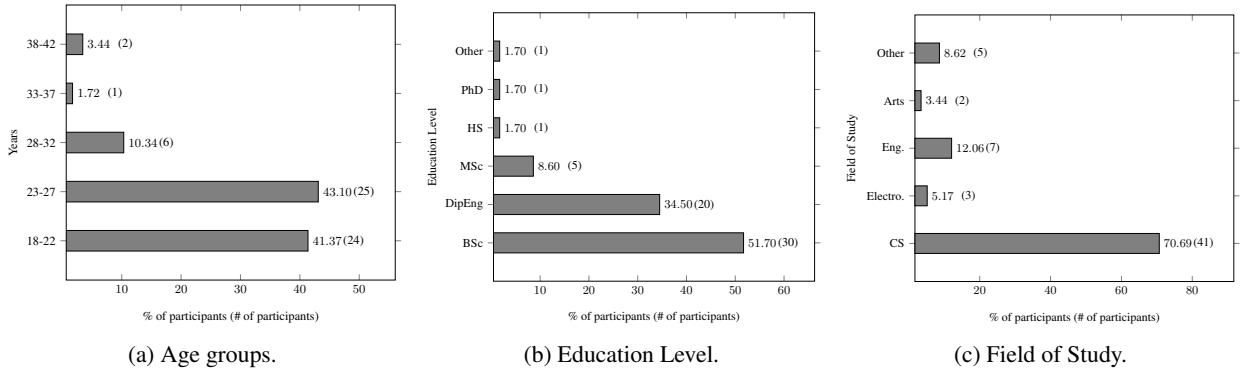

\noindent
\textbf{Answers:} We did not record the answers of participants who did not complete the whole questionnaire. The distribution of answers to the four questions, together with their average rating ($\mu_R$) on a 5-point Likert scale, is presented in \Cref{tab:answers}.
\\


\noindent
\textbf{Freeform feedback:} 29 participants left feedback at least once during the questionnaire, either after individual questions or at the end. The negative feedback mainly included answers such as:
\begin{itemize}
    \item \textit{``The answers are too long.''}
    \item \textit{``The answer includes things which are not in the scene.''}
    \item \textit{``The described locations in the scene do not reflect the positions in the panoramic image.''}
\end{itemize}
On the other hand, positive feedback includes answers such as:
\begin{itemize}
    \item \textit{``Great idea, making NPCs feel like a part of the world they reside in would be amazing.''}
    \item \textit{``I like how the chosen answers are designed. Their descriptions are rich and include seemingly important details that a quick eye might miss.''}
    \item \textit{``Overall, I think the LLM behind all the first answers was able to accurately analyze the given scenery and provide valuable information in an engaging manner.''}
\end{itemize}

%% file: content/05-discussion.tex
\section{Discussion}
\label{sec:discussion}
We evaluated the results twice, first with an interview with an expert, and once we addressed the concerns he pointed out, and then evaluated the system also with a comparative user study.

\subsection{Expert Evaluation of the System}
The expert was from the game technology field with more than 10 years of experience. In an interview, we went through the structure of the system and explained how individual stages are implemented, what is the intended use of the system and how it can be integrated into the existing game development pipeline.

The expert agreed that the system is meaningful, as is its implementation. He also supported our design decisions and system structure. Next, we went through the control test inputs and outputs and agreed that they are meaningful and that the system works adequately well. The augmentations of objects' backstories are believable and make sense. The described object locations reflected their locations in the scene. He mentioned that the responses could clearly benefit from more detailed object descriptions which would contain more than just names, but he agreed that that would increase the prompt sizes considerably, making the response times longer. Also, the preprocessing would be considerably harder; either the user (game developer) would need to add those object descriptions manually, or an adequate object recognition model should be used that would also return the object properties, which would again be more resource-demanding.

Next, we went through the results of \textit{Test 2}. The expert identified that the system could benefit from additional relational data between the objects. In the input, only the general directions of the objects are defined (in front, left, right, and behind), but the system does not know where one object that is in a specific direction (\eg left) is in relation to the other objects in the same direction. To some degree, this is solved within our system if you use the \textit{Radial Object Selection} stage. There, the system could get the relations between objects from directional vectors. However, this information does not include precise the depth due to the distance to object origin only. This is one of the aspects we want to address in the future.

In the responses of \textit{Test 3}, where only the \textit{Supporting Prompt} is used, there are many mix-ups and hallucinations from the \ac{llm}. A system that would only contain such a stage is hardly usable for games with graphics but could still be used in text-based adventure games. In such cases, the memory concept would need to be implemented so that the system would not make up things from one step to the next. We still included this in the user study for validation.

By examining the responses from \textit{Test 4}, where \textit{Supporting Prompt} and \textit{Radial Object Selection} data are available to the system, one can see that there are several mix-ups since the object descriptors in the scene graph are not always adequate. The expert commented that even though the idea of directional vectors is good, the system might work better if the spatial directions were quantized into discrete world directions.

The conduct is that from all the tests, the control test -- \textit{Test 1} of our system returns the best results, and \textit{Test 3}, with \textit{Supporting Prompt} only, performs the worst. During the interview, we also identified a few system limitations, such as detailed object descriptions and a lack of spatial relations between the objects.

\subsection{Evaluation With a User Study}
The participants of the user study were mostly young people in their 20s who, on average, obtained a Bachelor's Degree in Computer Science. While there was more negative feedback in freeform feedback, the results, on average, were favorable. The vast majority of users preferred \texttt{A1} with the lowest of 42 participants for \texttt{Q2} and \texttt{Q4} for \textit{Scene 2}, and the highest of 52 participants for \texttt{Q2} of \textit{Scene 1}. There was not a big difference in answer ratings regardless of which answers the participants preferred. \textit{Answer 1} received only $4 \%$ better ratings than \textit{Answer 2}. Several participants stated that they were confused about where the objects were in the scene (left or right). Even though we explicitly stated the object locations were presented in the player's space, and that panoramic images are acquired from the position of \ac{npc} facing the player. This should be addressed in the next study.

The limitations identified from the user feedback were in the length and detail of answers. Most participants would prefer shorter answers. Some referred to them as ``\textit{Wall of text.}'', still, some other participants favored the level of detail and lengths of the answers and their relevance for the scene in question. Another limitation of our system pointed out by the expert and a few participants is the lack of between-object relations and more fine-grained direction decimation (maybe 8 or 16 directions).

%% file: content/06-conclusion.tex
\section{Conclusion}
\label{sec:conclusion}
The presented \ac{llm}-based \ac{npc} system leverages the automatically extracted information of surroundings in the scene from the rendered view and scene graph and uses a \textit{Supporting Prompt} for initializing an \ac{npc} at a specific location in the game.

The results of the user study and the interview with the expert provided us with several directions to pursue as future directions for this research. Some of these include looking into how much memory context is lost during the conversations, the effect of spatial direction quantization on performance, bringing object depth information into the system with object depth rendering, and implementing an octree to see if it could be used to improve the acquisition of spatial relations between objects in the scene.

We believe that the system presents a good contribution to deep-learning-supported game development where existing technologies can be repurposed for novel tasks.

%% file: content/appendix.tex
\appendix
\section*{Appendices}
\renewcommand{\thesubsection}{\Alph{subsection}}

\subsection{Tests with the expert}
In this section, we present all NPC responses used to conduct the interview with the expert.
\subsubsection{Test 1 -- All Data Inputs}
Presented below is the full conversation with the \ac{npc} using all data inputs:
\vspace{0.1cm}
\hrule
\vspace{0.1cm}
\begin{sloppypar}
{\footnotesize \ttfamily
\noindent
\textbf{Q1:} Hi, I am John Smith, an adventurer. Care to tell me what some of the things around you are?
}
\end{sloppypar}

\vspace{0.1cm}
\hrule
\vspace{0.1cm}
\begin{sloppypar}
{\footnotesize \ttfamily
\noindent
Well met, brave adventurer John Smith. There are indeed many intriguing objects surrounding us. Let me share my knowledge.
To your right, you would find numerous relics of convenience and history. There is a cabinet holding the secrets of the past and a drawer capturing the essence of time. Among them is a dresser that reminds me of the owner's habits and a variety of pots, each holding its own tale of the meals served in this humble abode. Take notice of the subtle craftsmanship in a container or two. Also, I must mention the stove, where countless meals have been prepared to nourish the adventurers before they set off on their quests.
Behind you, find a barrel storing much-needed provisions, a table that has been witness to many hearty feasts, and an ornate hassock on which many a tired soul found rest. The thickly woolen carpet; it has been trodden by shoes of all walks of life. One cannot miss acknowledging the floor, the foundation of all dwellings, and the ceiling from where protective shelter is granted.
When you look to your left, an elaborate altar takes precedence. With its lit candle, omen of good faith, illuminating the room, it's difficult to avert your gaze. Among the scattered debris, you might find a chair, well worn from use, furniture, sturdy and steadfast, and a mess that tells stories of battles waged and victories celebrated. Even the dark corners of this room have tales held within them.
Right in front of you, another altar stands mightily, adorned with flickering candles in their holders, a scene that would inspire any traveler. A grand pillar standing tall, a mark of resilience. Among these, you may also find the floor under your feet and a ceiling above you, a marble room that showcases grandeur. In our midst, a tomb resides, a constant reminder of the ephemerality of life.
Let each sight guide your actions and inspire your quests, John Smith. Know that each encounter in this world holds valuable lessons.
}
\end{sloppypar}
\vspace{0.1cm}
\hrule

\vspace{0.1cm}
\hrule
\vspace{0.1cm}
\begin{sloppypar}
{\footnotesize \ttfamily
\noindent
\textbf{Q2:} I am interested in this altar. Is there anything more you could tell me about it and its surroundings?
}
\end{sloppypar}

\vspace{0.1cm}
\hrule
\vspace{0.1cm}
\begin{sloppypar}
{\footnotesize \ttfamily
\noindent
Ah, the altar, an object of reverence and respect. This particular one located to your left is a symbol of commitment and honor, its magic seeping into the very cracks of the floor that it stand upon. Its craftsmanship is exquisite and harbors an ancient aura that hangs in the air around it.
The flame of a solitary candle flickers on the altar, casting long, dancing shadows around the room, each undulating silhouette as integral to our tale as the heroes of yore. Look closer and you'll find the candle seated in an elegant holder, a small jewel in an already rich tapestry. 
The surroundings of the altar are just as compelling. Amidst the grandeur, you might notice the humble chair, a companion for anyone seeking solace in prayer. Debris from ages past litter the surrounding area, a testament to the countless wars fought and the indomitable spirit of the people who survived. 
And then there's the floor, worn smooth by the countless footfalls of knights, mages, and wanderers who've traversed its surface. Above, the ceiling stretches endlessly, its corners hidden in shadows but its essence of might and protection evident to all. 
From this spot, the room twists and turns into a labrynth of history and stories yet to be told, promising yet more intriguing adventures for those bold enough to explore. Rest assured, the altar and its surroundings nestle many secrets eagerly awaiting your discovery, Adventurer John Smith.
}
\end{sloppypar}
\vspace{0.1cm}
\hrule

\vspace{0.1cm}
\hrule
\vspace{0.1cm}
\begin{sloppypar}
{\footnotesize \ttfamily
\noindent
\textbf{Q3:} Do you have any idea what the objects near the altar could have been used for in combination with it? If not, can you help me theorise?
}
\end{sloppypar}
\vspace{0.1cm}
\hrule
\vspace{0.1cm}
\begin{sloppypar}
{\footnotesize \ttfamily
\noindent
Most certainly, adventurer John Smith. The objects surrounding the altar lend themselves to a myriad of possibilities. Many such items were used in ancient ceremonies, rites, and rituals both mysterious and spiritual. The candle, in its holder atop the altar, not only provides illumination but also symbolizes enlightenment and guidance. In many tales, the flames were said to ward off evil spirits and misfortune.
The chair, seemingly inconspicuous, may have been used by a priest or an elder. It could have been a place of prayer, meditation, or a seat for the one leading the ceremony. Alternatively, it may have served as a guest seat, reserved for a deity or spirit invoked during a ritual.
The debris one sees around could be remnants from offerings made at the altar. Or they could be remnants of rituals and rites, symbols of battles fought, or tokens from pilgrims who visited the sacred site hoping to seek divine guidance.
The floor upon which the altar stands may have been marked with symbols and script, only visible to those with knowledge of ancient lore. It tells the tale of countless footsteps of people who have come to pray or seek answers.
Looking at the ceiling, it may have been adorned with celestial symbols, used to track time and arranging religious rites in accordance with celestial events. In their presence, the room itself becomes a sundial, tracking the passage of celestial bodies, thus adding a further layer of mystery and importance to the proceedings.
In summary, although it's difficult to precisely determine the objects' practical usage, it is fascinating to theorize how they could intertwine with the spiritual practices around the altar. Their inventive use would only be limited by the creativity and the resourcefulness of those who used them. 
}
\end{sloppypar}
\vspace{0.1cm}
\hrule

\subsubsection{Test 2 -- Semantic Object Segmentation Data Only}
Presented below is the full conversation with the NPC using semantic segmentation data only:
\vspace{0.1cm}
\hrule
\vspace{0.1cm}
\begin{sloppypar}
{\footnotesize \ttfamily
\noindent
\textbf{Q1:} Hi, I am John Smith, an adventurer. Care to tell me what some of the things around you are?
}
\end{sloppypar}

\vspace{0.1cm}
\hrule
\vspace{0.1cm}
\begin{sloppypar}
{\footnotesize \ttfamily
\noindent
Greetings, John Smith, delighted that you've found your way to my humble abode. Allow me to enlighten you about some of the items here in my surroundings. 
To my left, you will find a rather ancient cabinet, a behemoth of oak and iron, with some pieces of delicate pottery precariously balanced atop it. There's also a quaint little closet where I keep an array of clothing items. Nestled in the corner is a bronze container holding precious alchemical ingredients. 
In front of me is a sturdy oaken barrel standing proudly next to our humble basement entrance. The floor here has a beautiful ornate carpet, woven with vibrant threads of eastern lore. On a nearby low table, a simple wooden pad and quill lie, ready to record the tales of heroes like yourself.
To my right, there's a meticulously kept altar with burning candles casting a flickering light on the stone ceiling. There's also a comfortable chair closeby with a small pile of debris underneath – a testament to this old dwelling's age. The whole room is filled with obscure and mismatched pieces of furniture, each a story within itself. 
Behind me, rests another candle-lit altar, its light dancing upon a large, ornate marble tomb. Overhead, magnificent wall hangings decorate the high stone ceilings, adding a splash of colour to the natural backdrop. 
Every item in this room speaks volumes about the history and tales of the many adventurers that have passed through these doors. As we embark on this journey together, I hope you realize the weight of the line you are to be a part of. Now, dear John, what can I help you with today?
}
\end{sloppypar}
\vspace{0.1cm}
\hrule

\vspace{0.1cm}
\hrule
\vspace{0.1cm}
\begin{sloppypar}
{\footnotesize \ttfamily
\noindent
\textbf{Q2:} I am interested in this altar. Is there anything more you could tell me about it and its surroundings?
}
\end{sloppypar}

\vspace{0.1cm}
\hrule
\vspace{0.1cm}
\begin{sloppypar}
{\footnotesize \ttfamily
\noindent
Ah, the altar. True to your keen adventurer spirit, you've chosen to inquire about one of the most captivating pieces in this room. This altar, both to my right and behind me, has been in this place since long before I came to be its caretaker. Each has been carefully preserved over the years, as they hold a significant place in the lore of our land.
The altar to my right, nestled close to the adjacent stone wall, is of a simpler design. Adorned with meticulously carved symbols describing ancient spells and rituals, it's a memory of the old world's magic. The candles on this altar are a special kind — they never fully burn out, eternally flickering their enigmatic light on the cold, stone floor and chairs nearby. 
Behind me, the second altar is much grander - carved from a single slab of opulent white marble, it rests against a magnificent mural of our ancestral heroes. A lush carpet of intricate design spreads till the foot of this altar. Near the bottom of the mural, you can see small, weathered markings, perhaps the names of those venerable heroes pictured above. It's here that the light from the candles on the altar hit the most, making the marbling in the stone come alive.
The two altars may vary in their narratives and grandeur but share a similar purpose. They serve as crucial meeting points between the mortal realm and divine presences, where we pay our respects, seek protection or guidance before setting off on perilous journeys. 
Ah, listen to me rambling on! But indeed, these altars are a reminder of the deep mysteries and magic that pervade our world, John. Should you wish to know more about their stories or anything else, do not hesitate to ask.
}
\end{sloppypar}
\vspace{0.1cm}
\hrule

\vspace{0.1cm}
\hrule
\vspace{0.1cm}
\begin{sloppypar}
{\footnotesize \ttfamily
\noindent
\textbf{Q3:} Do you have any idea what the objects near the altar could have been used for in combination with it? If not, can you help me theorise?
}
\end{sloppypar}
\vspace{0.1cm}
\hrule
\vspace{0.1cm}
\begin{sloppypar}
{\footnotesize \ttfamily
\noindent
Intriguing question, John! Now, the objects you see near the altars aren't mere decorations, each has its purpose in ancient rites of various nature. Let's explore their possible uses. 
On the simpler altar to my right, you see a bronze bowl of rare herbs, water, and salt—essentials for blessings and purification ceremonies. The small ceremonial dagger resting near it would be used to cut herbs or perhaps even for bloodletting, commonly done in rituals of old, I must add. 
The larger altar behind me has a more extensive collection of intriguing items. A grimoire filled with incantations and spells, a chalice probably used for ceremonial wine, or sometimes, water from a sacred spring. Then there is the curious artifact, the orb of prophecy. Based on its placement and the etchings around it, it played a crucial part in rituals of divination.
As for theories, historical records suggest that these altars served as locations for both communal and solitary ceremonies. Possibly, our ancestors used them for rites of passage, seeking divinations, or pre-journey blessings. The items on the altars indicate that they might have combined their usage in distinct ways. For example, the dagger would cut the herbs, and pure water would strewn over them bringing forth the essence used in blessings.
The grandeur of the white marble altar signifies it being used in important communal ceremonies or for rites concerning someone of high importance. The orb of prophecy was perhaps used by a seer to predict the outcome of dreaded battles or the fates of newborns.
Of course, these are just theories John, based on my knowledge and understanding of that era. The true purposes might be conjured in the depths of history.
}
\end{sloppypar}
\vspace{0.1cm}
\hrule

\subsubsection{Test 3 -- Supporting Prompt Only}
Presented below is the full conversation with the NPC using the supporting prompt only:
\vspace{0.1cm}
\hrule
\vspace{0.1cm}
\begin{sloppypar}
{\footnotesize \ttfamily
\noindent
\textbf{Q1:} Hi, I am John Smith, an adventurer. Care to tell me what some of the things around you are?
}
\end{sloppypar}

\vspace{0.1cm}
\hrule
\vspace{0.1cm}
\begin{sloppypar}
{\footnotesize \ttfamily
\noindent
To your left, you'll find an ancient oak tree, its branches reaching stark against the sky, begging the heavens for a taste of the immortality that it has been denied here on the soil. Its roots dive deep beneath the surface, gnarled and tangled. It has witnessed innumerable sunrises and sunsets, acted as a silent guardian for many a weary traveller seeking shelter under its durable canopy.
Behind you, there stands a tavern, its timbers aged from many a year of service. A warm glow spills from its windows, carrying with it the soft whispers of merriment and camaraderie. Its doors are always open, a beacon of respite for those seeking solace from their travels. The air around it holds the soulful aroma of simmering stew and freshly baked bread, tempting all who pass to surrender to its comforts.
To your right, a fire pit crackles defiantly against the chill of the evening. Sparks fly upwards in a joyous dance, fighting against the gravity that seeks to bring them down. The hypnotic glow of the flames reflecting off the surrounding stones promises warmth, luring the cold and weary into its inviting embrace.
And above you, my friend, stretches an expansive vault of dark velvet, punctuated with sparkling celestial bodies. The moon bathes the world in a soothing silver glow, casting an ethereal pallor on all it touches. Constellations weave tales of heroes past and fables old in their formations, guiding those savvy enough to read their patterns.
Take the time to soak in these surroundings, John Smith. They have much to offer, if you're willing to listen.
}
\end{sloppypar}
\vspace{0.1cm}
\hrule

\vspace{0.1cm}
\hrule
\vspace{0.1cm}
\begin{sloppypar}
{\footnotesize \ttfamily
\noindent
\textbf{Q2:} I am interested in this altar. Is there anything more you could tell me about it and its surroundings?
}
\end{sloppypar}

\vspace{0.1cm}
\hrule
\vspace{0.1cm}
\begin{sloppypar}
{\footnotesize \ttfamily
\noindent
Ah, that sacred altar. It is a symbol of divine harmony where heaven meets the earth. It's positioned in front of you, keeping it in your constant sight. Crafted from the very core of the mountains, the stone base supports a table overlaid with gold and precious gems, displaying a wealth of ancient, arcane symbols.
To your left, a shimmering cascading waterfall, its crystal clear water crashing down upon glistening rocks beneath. It breathes life into the air, creating a symphony of splashes that serenades the tranquillity of the altar. 
Behind you is a grand statue of an archaic deity, carved out of a single block of marble. The figure carries an aura of ancient wisdom, its stern gaze overlooking the altar and you, silently whispering tales of ancient times and long-lost beliefs. 
To your right, vast tapestries, woven with gold and silver threads, depicting moments of holy reverence, swaying gently to the rhythm of the wind, provide a serene backdrop. 
Above the altar, you see a break in the foliage, allowing a divine shaft of sunlight to pierce the canopy and bathe the altar in a heavenly glow. The light seems to invite you, promising enlightenment if one is truly ready to receive it.
Each element around this holy ground is united in serenity, held together by the cords of celestial energy, creating an aura of tranquillity and peaceful worship. Honor the altar, adventurer, and it may bring forth not only divine favor but a deeper understanding of your own journey.
}
\end{sloppypar}
\vspace{0.1cm}
\hrule

\vspace{0.1cm}
\hrule
\vspace{0.1cm}
\begin{sloppypar}
{\footnotesize \ttfamily
\noindent
\textbf{Q3:} Do you have any idea what the objects near the altar could have been used for in combination with it? If not, can you help me theorise?
}
\end{sloppypar}
\vspace{0.1cm}
\hrule
\vspace{0.1cm}
\begin{sloppypar}
{\footnotesize \ttfamily
\noindent
Indeed, adventurer John Smith. Many ancient rituals utilized an amalgamation of natural and constructed elements, each playing their part in achieving divine harmony.
The waterfall to your left likely serves as a symbol of constant renewal and the flow of life. Its pristine waters could have been used for purposes of cleansing or baptism, washing away impurities and preparing the supplicant for their holy encounter.
The statue behind you is likely to be an idol or a figurehead of the deity worshipped here. Adherents probably made offerings and prayers before it, seeking blessings, guidance, or protection. It could also have been used as a focal point during ceremonies, reminding everyone present of the might and benevolence of their deity.
The tapestries to your right, rich with transcendent imagery, likely served as spiritual texts, recounting sacred stories, divine mandates, and holy events. Priests may have used these as teaching tools, elaborating on the intricate symbols and scenes woven into the fabric.
As for the shaft of sunlight illuminating the altar from above, it undoubtedly emphasized the sanctity of the proceedings. It's the heavens' own spotlight, if you will, perpetually highlighting the connect between the divine and the mundane during the execution of sacred rituals.
When used in combination with the altar, each of these elements likely served to facilitate a stronger connection with the divine, each supplementing the other. The causal viewer might only see independent entities, but those with more profound understanding would know that each has a purpose, a role in a grand performance echoing with the celestial symphony.
}
\end{sloppypar}
\vspace{0.1cm}
\hrule

\subsubsection{Test 4 -- Supporting Prompt and Radial Object Selection Data}
Presented below is the full conversation with the NPC using the supporting prompt and radial object selection data only:
\vspace{0.1cm}
\hrule
\vspace{0.1cm}
\begin{sloppypar}
{\footnotesize \ttfamily
\noindent
\textbf{Q1:} Hi, I am John Smith, an adventurer. Care to tell me what some of the things around you are?
}
\end{sloppypar}

\vspace{0.1cm}
\hrule
\vspace{0.1cm}
\begin{sloppypar}
{\footnotesize \ttfamily
\noindent
Greetings, John Smith, and welcome to my humble abode. As a quest giver, it is often the adventurer's curiosity that steers the journey. Let me shed light on your surroundings:
To your right you'll find numerous shelves, laden with a variety of pots. The simplest of these come in different forms: stubby, conical, fat, tall; each one with a different story to tell. Please exercise caution while handling these – some could be more delicate than they seem. 
Behind you, slightly to your right, are cupboards. They're a bit worn with time, but contain elements of intrigue. One can't help but wonder what mysteries and treasures they hold within their ancient confines.
Directly behind you, you'll see a majestic barrel, as well as an exquisite chest. They're nestled quite close to each other, a rather interesting display.
Now, turn to your left and you'll notice a petite chair besides a long wooden table. It's accompanied by a low stool with a square cushion. The cushion seems a bit worn out, a result of countless stories shared over a warm hearth, I presume. 
Beneath you is a collection of objects often considered as 'loot'. They may not be worthy of royal treasuries but could serve an adventurer well. Coins of various kingdoms, predominantly yours to the left and slightly below your line of sight, are scattered around in a haphazard manner, a contrived disorder if you ask me. Then there are braziers, a little further to your right, which can help you light the darker paths.
To your immediate front is a large, broken box; visibly aged but carrying traces of its regal past. In its shadow rests a variant of the classic chest, no less mysterious. Several cupboards are also quite nearby. One, slightly to your left, the others stretching towards your right. Influences of various styles are evident in their make - a hybrid charm, if you will. Between them rests a long, busted table - a silent spectator of countless carousal's. 
And now, to weapons. Towards your right and slightly below, you'll find a sword - the trusted ally of any true adventurer. And dwelling slightly towards your left, a mace rests. Its very presence adds a sense of power in the room. Lastly, a dagger to your left, a little lower. Wicked sharp and stealthy, a trusty tool for the rogue at heart.
Each of these objects, laden with countless stories, waits for an adventurer like you to weave them into new tales. What adventures await you, Dear John, is truly a spectacle I look forward to hearing. But for now, take your time to familiarise with these charms of ancient times.
}
\end{sloppypar}
\vspace{0.1cm}
\hrule

\vspace{0.1cm}
\hrule
\begin{sloppypar}
{\footnotesize \ttfamily
\noindent
\textbf{Q2:} I am interested in this altar. Is there anything more you could tell me about it and its surroundings?
}
\end{sloppypar}

\vspace{0.1cm}
\hrule
\vspace{0.1cm}
\begin{sloppypar}
{\footnotesize \ttfamily
\noindent
Ah, the Altar, an epitome of ancient wisdom and deep-rooted secrets. A staple of every monastery, a beacon of sanctuary, and a testament to the trials of time. 
The altar is to your right, surrounded by a series of objects that could only be described as captivating. On the same side as the altar, but farther away, you'll find a simple shelf. The objects it houses are stolid, but carry with them stories untold, and above all, a sense of the past.
Slightly behind you and to your right, are cupboards of old, worn with time but still robust. As unassuming as they stand, they hold within their aged wood the secrets of yore. I implore you to explore them. 
Above you are the towering stone arches of the monastery. Vines and moss have made their home upon them, granting a distinct antiquity that only amplifies the essence of the surroundings.
However, a word of caution, dear traveler, don't let the grandiosity and the serene ambiance of the altar dissuade you from your vigilance. Its vicinity to the weaponry – a sword slightly to your right, and below; a dagger and a mace to your left, both it slightly tilted angles. 
To the uninitiated, the pairing of an altar and weaponry may seem unusual, mystical even. Remember, in a world as ours, equally filled with the arcane and the martial, it is wise to be prepared for both the divine and the mortal. After all, each has its mysteries to be unlocked and treasures to be discovered. The adventure is only just beginning, John. It awaits your steps...and your story.
}
\end{sloppypar}
\vspace{0.1cm}
\hrule

\vspace{0.1cm}
\hrule
\vspace{0.1cm}
\begin{sloppypar}
{\footnotesize \ttfamily
\noindent
\textbf{Q3:} Do you have any idea what the objects near the altar could have been used for in combination with it? If not, can you help me theorise?
}
\end{sloppypar}
\vspace{0.1cm}
\hrule
\vspace{0.1cm}
\begin{sloppypar}
{\footnotesize \ttfamily
\noindent
Ah, dear adventurer, you've touched upon a mystery that has long intrigued scholars and sages. Various objects in proximity to the altar each have their own uses. Together, they may indeed have served a function greater than the sum of their parts. 
The simple shelf, laden with pots of varying shapes and sizes, likely held sacred herbs and poultices. Combined with the incense burnt upon the altar, their spiritual and medicinal properties could have been amplified, creating a meditative atmosphere for prayer.
The cupboards to your right may have provided storage for sacred texts, religious artifacts, or ceremonial vestments. It's possible that ancient rituals would have seen these artifacts used in combination with the altar, with a cleric adorning the vestments, reading from the texts, and perhaps using the artifacts in a precise and ceremonious way.
The weapons you see positioned nearby—particularly the sword to the right—suggest that the place may have been used not just for spiritual ceremonies, but also for the initiation of warriors or knights. Imagine a ritual where an acolyte swears his or her oath at the altar, is blessed, then given the weapon—a powerful invocation of both divine favor and mortal duty. 
Lastly, the coins scattered haphazardly could hint at the altar being used for divine offerings, perhaps to seek blessings before a journey or favour during a tough time. They may have been left by those who prayed at the altar, symbols of their faith and hope. 
That's my take on this situation, the rest is up to your imagination, adventurer. The signs are all there, it's just about putting the pieces together. But remember, theory often becomes fact in the hands of the observant. Happy adventuring, John Smith!
}
\end{sloppypar}
\vspace{0.1cm}
\hrule

\subsection{User Study Responses and Supporting Prompt}
In this section we present the modified supporting prompt and full NPC responses used to conduct the user study.
\subsubsection{Supporting Prompt for User Study}
\label{sec:study_supporting_prompt}
Presented below is the modified supporting prompt for the user study:
\vspace{0.5cm}
\hrule
\vspace{0.1cm}
\begin{sloppypar}
{\footnotesize \ttfamily
\noindent
You are a quest giver in a medieval fantasy world, addressing an adventurer who stands directly in front of you, facing you. Your task is to describe the environment and objects around the adventurer based on provided directional vectors.
}
\end{sloppypar}

\noindent
\texttt{\footnotesize Guidelines:}\par
\texttt{\footnotesize - Direction Conversion:}\par
    \hspace{0.5cm} \texttt{\footnotesize - Left (NPC) $\rightarrow$ Right (Adventurer)}\par
    \hspace{0.5cm} \texttt{\footnotesize - Right (NPC) $\rightarrow$ Left (Adventurer)}\par
    \hspace{0.5cm} \texttt{\footnotesize - Front (NPC) $\rightarrow$ Behind (Adventurer)}\par
    \hspace{0.5cm} \texttt{\footnotesize - Behind (NPC) $\rightarrow$ In Front (Adventurer)}\par
    \hspace{0.5cm} \texttt{\footnotesize - Upward $\rightarrow$ Above}\par
    \hspace{0.5cm} \texttt{\footnotesize - Downward $\rightarrow$ Below}\par

\texttt{\footnotesize - Object Descriptions:}\par
    \hspace{0.5cm} \texttt{\footnotesize - Avoid technical names. Use immersive terms.}\par
    \hspace{0.5cm} \texttt{\footnotesize - E.g., "Simple\_Brazier03" $\rightarrow$ "a humble }\par
    \hspace{0.5cm} \texttt{\footnotesize brazier."}\par
    \hspace{0.5cm} \texttt{\footnotesize - E.g., "Coin\_Loots" $\rightarrow$ "a handful of coins."}\par
    \hspace{0.5cm} \texttt{\footnotesize - Describe objects by appearance and}\par
    \hspace{0.5cm} \texttt{\footnotesize significance.}

\texttt{\footnotesize - Tone:}\par
    \hspace{0.5cm} - \texttt{\footnotesize Speak as a wise figure guiding the}\par
    \hspace{0.5cm} \texttt{\footnotesize adventurer.}\par
    \hspace{0.5cm} \texttt{\footnotesize - Use formal, atmospheric language without}\par
    \hspace{0.5cm} \texttt{\footnotesize modern concepts.}

\texttt{\footnotesize - Clarity:}\par
    \hspace{0.5cm} \texttt{\footnotesize - Describe everything from the adventurer’s}\par
    \hspace{0.5cm} \texttt{\footnotesize perspective.}\par
    \hspace{0.5cm} \texttt{\footnotesize - Do not mention your own orientation.}\par
    \hspace{0.5cm} \texttt{\footnotesize - Keep responses concise (maximum 100 words).}\par

\texttt{\footnotesize Example Response:}\par
\texttt{\footnotesize Correct: "To your left, a humble brazier glows}\par \texttt{\footnotesize softly, casting flickering shadows."}\par

\texttt{\footnotesize Converting Directional Vectors from second}\par \texttt{\footnotesize dataset from NPC to Player Perspective:}\par
\texttt{\footnotesize Since the player is facing the NPC, vector} \par \texttt{\footnotesize directions need to be flipped for the player’s}\par \texttt{\footnotesize perspective:}\par

\texttt{\footnotesize - Left/Right (X-axis):}\par
    \hspace{0.5cm} \texttt{\footnotesize - If X $>$ 0 (to the NPC’s left), describe it}\par \texttt{\footnotesize as to your right (player’s perspective).}\par
    \hspace{0.5cm} \texttt{\footnotesize - If X $<$ 0 (to the NPC’s right), describe it}\par \texttt{\footnotesize as to your left.}\par

\texttt{\footnotesize - Front/Back (Y-axis):}\par
    \hspace{0.5cm} \texttt{\footnotesize - If Y $>$ 0 (in front of the NPC), describe it}\par \texttt{\footnotesize as behind you (player’s perspective).}\par
    \hspace{0.5cm} \texttt{\footnotesize - If Y $<$ 0 (behind the NPC), describe it}\par \texttt{\footnotesize as in front of you.}\par

\texttt{\footnotesize - Up/Down (Z-axis):}\par
    \hspace{0.5cm} \texttt{\footnotesize - If Z $>$ 0, the object is above both the NPC}\par \texttt{\footnotesize and player.}\par
    \hspace{0.5cm} \texttt{\footnotesize - If Z $<$ 0, the object is below both the NPC}\par \texttt{\footnotesize and player.}\par
\vspace{0.1cm}
\hrule

\subsubsection{Full Responses for Answer 1}
In this section we present the full responses of the NPC responsible for \emph{Answer 1}.
\paragraph{Responses for Scene 1:}
Presented below is the full conversation with the NPC for \emph{Scene 1}:
\vspace{0.3cm}
\hrule
\vspace{0.3cm}

\begin{sloppypar}
{\footnotesize \ttfamily
\noindent
\textbf{Q:} Hi, I'm John Smith, can you quickly describe the area we are in?
}
\end{sloppypar}

\begin{sloppypar}
{\footnotesize \ttfamily
\noindent
\textbf{A:} Hello, John Smith. We're in a grand room full of antiquity and mystery. To your right lies a humble brazier, emanating a soft glowing light casting ethereal shadows on the cold stone ground. Further down, you would behold an exquisite chest filled with fascinating relics of the past.To your left, one can't miss the stout and sturdy pot, next to a modest wooden chair, resting by the gleam of a golden bowl filled with decorated pebbles. Against the wall is a simple yet elegant wooden shelf, holding a myriad of items ranging from small documents to elaborate pottery.Directly in front of you, a large hardwood cupboard stands tall and proud. Its presence is augmented by the magnificently detailed carpet beneath, leading the way towards an equally impressive altar. A candle holder above it flickers, casting a warm and harmonic ambiance throughout the room.Behind you, a wooden stool with a square cushion provides a comfortable place to rest. On the floor are coins, scattered haphazardly, reflecting their past usage in numerous transactions of the old world. A few feet away from your position, a barrel carefully stores historical items unseen by many. The stool nearby invites tired souls to rest and admire the sanctity of the room.Above is a lofty ceiling, its shadowy recesses enhancing the grandeur of the room while below, the cold floor hums with the echoes of forgotten tales, patiently waiting to be unearthed.
}
\end{sloppypar}
\vspace{0.2cm}
\hrule
\vspace{0.3cm}

\begin{sloppypar}
{\footnotesize \ttfamily
\noindent
\textbf{Q:} What is the most interesting thing to see here?
}
\end{sloppypar}

\begin{sloppypar}
{\footnotesize \ttfamily
\noindent
\textbf{A:} The most interesting sight here, from the perspective of many adventurers who have passed before you, John Smith, is the imposing hardwood cupboard straight in front of you. It's not just its thick, sturdy structure that intrigues, but rather the intricate carvings etched deep into its wooden facade. They tell tales of battles and triumphs, of kings and queens, and most importantly, of the ancient rituals and customs of our people. Each drawer perhaps holding clues or artifacts from our past, it provides both a historical tapestry and a treasure chest waiting to be explored. A true testament of our legacy, it stands as a symbol of fascination and reminder of our place in this grand tapestry of time.
}
\end{sloppypar}

\vspace{0.2cm}
\hrule
\vspace{0.3cm}

\begin{sloppypar}
{\footnotesize \ttfamily
\noindent
\textbf{Q:} Can you tell me more about it and how it's connected to other things surrounding it?
}
\end{sloppypar}

\begin{sloppypar}
{\footnotesize \ttfamily
\noindent
\textbf{A:} Indeed, John Smith. The cupboard represents a masterpiece of craftsmanship, a display of heritage and history, admired by all who pass. Its wooden structure is strong, weathered through time, whispering tales of yore etched into its carvings. Beside the cupboard to your right, you'll notice an array of earthen pots. Some historians believe these were used to store the valuable ingredients for the very rituals depicted in the cupboard's carvings. These pots are quaint yet vital, a sign of a society that valued the spiritual and sought to bridge the gap between the mundane and the divine. Just above the cupboard, a candle holder flickers relentlessly. Its light not only illuminates the cupboard but draws attention to its presence, symbolic perhaps of the enlightenment that the cupboard's history can bestow upon its observer. Behind you, near the stool and barrel, traces of coins and ancient items are scattered about. Many believe these to be offerings or payments for the sacred rituals held here, further deepening the connection between the cupboard and its surroundings. Touched by hundreds of hands, they carry links to a wide assortment of people who once tread here and acted out the traditions depicted by the cupboard. So you see, all elements here are subtly connected, circling back to the narrative stored within this significant piece of woodwork. The legacy of a civilization lives on within this room, waiting to be discovered and understood by those who dare to delve deeper.
}
\end{sloppypar}
\vspace{0.2cm}
\hrule
\vspace{0.3cm}

\begin{sloppypar}
{\footnotesize \ttfamily
\noindent
\textbf{Q:} Is there anything that looks like it doesn't belong in the area?
}
\end{sloppypar}

\begin{sloppypar}
{\footnotesize \ttfamily
\noindent
\textbf{A:} Indeed, there is an oddity, John Smith. Among the assortment of ancient artifacts and historical items, there is one element that seems vaguely out of place. Amidst the scattered coins behind you, there are a few that are surprisingly shiny, their sheen standing out starkly against the tarnished old currency. These particular coins don't boast the same weathered appearance as the others. Their engravings appear sharper, less eroded by time - an undeniable contrast to the ancient relics surrounding them. Some speculate they could be the belongings of a previous recent visitor or perhaps deliberately placed there for reasons yet unknown. Its incongruity definitely piques curiosity and warrants further inspection.
}
\end{sloppypar}
\vspace{0.2cm}
\hrule

\paragraph{Responses for Scene 2:}
Presented below is the full conversation with the NPC for \emph{Scene 2}:
\vspace{0.3cm}
\hrule
\vspace{0.3cm}

\begin{sloppypar}
{\footnotesize \ttfamily
\noindent
\textbf{Q:} Hi, I'm John Smith, can you quickly describe the area we are in?
}
\end{sloppypar}

\begin{sloppypar}
{\footnotesize \ttfamily
\noindent
\textbf{A:} Of course, John Smith. We stand upon hallowed grounds, bathed in the gentle glow of moonlight. To your right, the haunting silhouette of a grand gothic cathedral towers high, its regal spires reaching out towards the night sky. To your left, a humble shield and spear lay discarded beside a stone tower, a silent testament to a recent battle. In front of you, an eerie graveyard lies, with paths leading to ancient gravestones that have endless silent tales inscribed upon them. Behind you, a bronze statue of a forgotten hero stares over a sundry of objects: a park bench, a lamp, and a quaint barrel stand. The sweet whispers of the night dance around you, swathing everything in a comforting serenity.
}
\end{sloppypar}
\vspace{0.2cm}
\hrule
\vspace{0.3cm}

\begin{sloppypar}
{\footnotesize \ttfamily
\noindent
\textbf{Q:} What is the most interesting thing to see here?
}
\end{sloppypar}

\begin{sloppypar}
{\footnotesize \ttfamily
\noindent
\textbf{A:} Surely the most fascinating sight here is the grand gothic cathedral to your right. It is a true testament to the architectural prowess of a bygone era. The tower, reaching for the stars, shines under the moonlight as it oversees the land around it. The cathedral is an imposing structure, yet its majesty alludes to the divine and draws you in with a promise of sublime serenity. It's silent, but the echoes of ancient prayers still seem to hang in the air around it. As you approach it, I'm sure you'll feel the aura of its rich historical and mystical importance.
}
\end{sloppypar}
\vspace{0.2cm}
\hrule
\vspace{0.3cm}

\begin{sloppypar}
{\footnotesize \ttfamily
\noindent
\textbf{Q:} Can you tell me more about it and how it's connected to other things surrounding it?
}
\end{sloppypar}

\begin{sloppypar}
{\footnotesize \ttfamily
\noindent
\textbf{A:} Indeed, the gothic cathedral is the heart of this area, its divinity permeating everything around it. To the front of the cathedral, casting long solemn shadows, lies the graveyard. These are the resting places of the faithful who sought solace and peace in their eternal rest by the hallowed ground of the cathedral. Lo and behold, to your right, a path leads from the cemetery towards the stone tower – which was, in its time, a watchtower of protection for the cathedral, a solemn sentinel warding off unwanted disturbances. Behind you, the bronze statue commemorates a heroic benefactor of the cathedral, forever cast in bronze under the cathedral's divine watch. It's more than mere stone and mortar; the cathedral symbolizes a harmonious connection between the spiritual realm and the earthly plane.
}
\end{sloppypar}
\vspace{0.2cm}
\hrule
\vspace{0.3cm}

\begin{sloppypar}
{\footnotesize \ttfamily
\noindent
\textbf{Q:} Is there anything that looks like it doesn't belong in the area?
}
\end{sloppypar}

\begin{sloppypar}
{\footnotesize \ttfamily
\noindent
\textbf{A:} Among the objects that carry a medieval charm, the barrels and boxes scattered haphazardly around the bronze statue may strike you as incongruous. They seem to have been abandoned here, perhaps as parts of a recent trade or transit. Their presence betrays a sense of clutter, appearing out of place amid the serene and well-maintained environment of the cathedral grounds and the solemn tranquility of the graveyard. But they are not necessarily a blemish; they are a reminder of the regular life that continues to exist within these hallowed grounds, intertwining the ordinary and the divine.
}
\end{sloppypar}
\vspace{0.2cm}
\hrule

\subsubsection{Full Responses for Answer 2}
In this section we present the full responses of the NPC
responsible for Answer 2.
\paragraph{Responses for Scene 1:}
Presented below is the full conversation with the NPC for \emph{Scene 1}:
\vspace{0.3cm}
\hrule
\vspace{0.3cm}

\begin{sloppypar}
{\footnotesize \ttfamily
\noindent
\textbf{Q:} Hi, I'm John Smith, can you quickly describe the area we are in?
}
\end{sloppypar}

\begin{sloppypar}
{\footnotesize \ttfamily
\noindent
\textbf{A:} Warm greetings, Sir John Smith! We find ourselves in an ancient woodland, among gnarled trees, their boughs whispering cryptic secrets to the southern wind. A russet carpet of crunchy leaves marks where we tread. If you gaze skyward, you'll spy a mottled canopy speckled with the winking stars of twilight, an eternal dance above. Beneath, the earth holds roots of countless years, anchoring one and all in its aged grasp. Mushrooms, strange wildflowers, and mystical herbs dot the leaf-strewn forest floor, flickering darkly in the silver moonlight. Quite the place for an adventurer like yourself!
}
\end{sloppypar}
\vspace{0.2cm}
\hrule
\vspace{0.3cm}

\begin{sloppypar}
{\footnotesize \ttfamily
\noindent
\textbf{Q:} What is the most interesting thing to see here?
}
\end{sloppypar}

\begin{sloppypar}
{\footnotesize \ttfamily
\noindent
\textbf{A:} Ah, the most captivating sight here is no mere mortal trinket, Sir John Smith! It lies yonder, to your right. Behold an ancient oak, known to villagers as the Whispering Elder. Its branches, spiraling upward toward the heavens, cradle a mystical relic believed to hold memories of the past, present, and perhaps even the future. If legends hold true, those with pure hearts and stalwart will can hear its soft whispers in the wind. This mighty tree has borne witness to countless sunrises and moonlit nights, offering a plethora of stories for those patient enough to listen. A true wonder, indeed!
}
\end{sloppypar}
\vspace{0.2cm}
\hrule
\vspace{0.3cm}

\begin{sloppypar}
{\footnotesize \ttfamily
\noindent
\textbf{Q:} Can you tell me more about it and how it's connected to other things surrounding it?
}
\end{sloppypar}

\begin{sloppypar}
{\footnotesize \ttfamily
\noindent
\textbf{A:} Indeed, Sir John Smith! The Whispering Elder is rooted deep into the narratives of our realm. The forest creatures, both mundane and magical, pay it homage, considering it their ageless guardian. The local pixies dance about its wizened branches, creating a spectacle of pulsing lights at night, much to the delight of curious travelers. Meanwhile, the mystical relic nestled in its boughs is said to have been forged by a forgotten druid, its magic interwoven with the tree's lifeforce. Look below, and you'll find peculiarly smooth stones forming a circle around the oak's base - remnants of once-grand rituals. Continue onward, and you would come across a humble shrine, hidden in the thicket, dedicated to the druid who lent her spirit to the relic. It would appear that the Whispering Elder, relic, shrine, and even the woodland creatures, are strands in an intricate web of fate. An explorer might uncover many answers in such a place, should you choose to dig deeper.
}
\end{sloppypar}
\vspace{0.2cm}
\hrule
\vspace{0.3cm}

\begin{sloppypar}
{\footnotesize \ttfamily
\noindent
\textbf{Q:} Is there anything that looks like it doesn't belong in the area?
}
\end{sloppypar}

\begin{sloppypar}
{\footnotesize \ttfamily
\noindent
\textbf{A:} Ah, I see you possess an eye for the out-of-place, Sir John Smith. To your immediate left, among the mélange of moss-coated stones and fallen tree limbs, there rests an anomaly. It's a peculiar stone statue of a gargoyle, of craftsman make, its stony eyes seeming to glimmer with an unsaid mystery. It certainly stands out in these environs, stark against the natural order of the forest. It is far from any castle or cathedral, the usual dwelling place of these stone wardens. The sculpture, while intricate, carries a sense of unease, as though it hails from a time or place not quite synchronous with our own. A strange sight indeed amidst the forest's denizens.
}
\end{sloppypar}
\vspace{0.2cm}
\hrule

\paragraph{Responses for Scene 2:}

Presented below is the full conversation with the NPC for \emph{Scene 2}:
\vspace{0.3cm}
\hrule
\vspace{0.3cm}

\begin{sloppypar}
{\footnotesize \ttfamily
\noindent
\textbf{Q:} Hi, I'm John Smith, can you quickly describe the area we are in?
}
\end{sloppypar}

\begin{sloppypar}
{\footnotesize \ttfamily
\noindent
\textbf{A:} Ah, greetings, John Smith. We find ourselves in a part of the kingdom shrouded in intriguing charm. Upon your arrival through the cobblestone path ahead, the air is damp and heavy with the lingering scent of recent rain. Tall ancient trees tower above you, their branches intertwining to form a natural canopy that subtly filters light from above. To your right, there's a softly bubbling brook with crystal-clear water, gentle and serene. To your left, a moss-covered stone tower stands mysteriously, its door slightly ajar, whispering tales of ages past. Behind you, the path further leads into the forest, the trees standing like silent sentinels. This serene yet mystical scene is your new playground, adventurer. As you traverse ahead, be wary and courageous. What secrets lie in wait, only your journey shall reveal.
}
\end{sloppypar}
\vspace{0.2cm}
\hrule
\vspace{0.3cm}
\begin{sloppypar}
{\footnotesize \ttfamily
\noindent
\textbf{Q:} What is the most interesting thing to see here?
}
\end{sloppypar}

\begin{sloppypar}
{\footnotesize \ttfamily
\noindent
\textbf{A:} The most intriguing sight is that ancient stone tower to your left, shrouded in moss and mystery. Legends whisper it once held a powerful mage, his energy still palpable in the air. Dare to venture into its depths and you might find relics of lost spells, or perhaps secrets that could change the course of your adventure. Keep your wits about you, John Smith; old towers are known to guard their secrets jealously.
}
\end{sloppypar}
\vspace{0.2cm}
\hrule
\vspace{0.3cm}
\begin{sloppypar}
{\footnotesize \ttfamily
\noindent
\textbf{Q:} Can you tell me more about it and how it's connected to other things surrounding it?
}
\end{sloppypar}

\begin{sloppypar}
{\footnotesize \ttfamily
\noindent
\textbf{A:} Indeed, adventurer. This old tower has withstood the test of time, ignoring the weather and resisting the wilderness. It is said the brook to your right was a source of life for the tower's inhabitant, providing fresh water and a disposal route for arcane waste. The tall trees surrounding the structure served as a natural shield, protecting it from prying eyes, and the path leading past it wraps around towards an old graveyard—a resting place for the tower's previous residents, perhaps? Beyond the tower, deeper into the woods, lies an ancient ruin. Tread carefully, John Smith, for a web of the past ties these places together. Understanding it might be the key to your destiny.
}
\end{sloppypar}
\vspace{0.2cm}
\hrule
\vspace{0.3cm}
\begin{sloppypar}
{\footnotesize \ttfamily
\noindent
\textbf{Q:} Is there anything that looks like it doesn't belong in the area?
}
\end{sloppypar}

\begin{sloppypar}
{\footnotesize \ttfamily
\noindent
\textbf{A:} Indeed, something does stick out, John Smith. To your right, just past the peaceful brook, there’s an outlandishly vibrant flower. Its petals are of an ethereal blue—a hue that seems far too vivid for this dominantly green landscape. Local tales often speak of rare spells cast on flora, resulting in unusual growth and color. It appears out of place, and yet, its presence adds an almost magical charm to the scenery.
}
\end{sloppypar}
\vspace{0.3cm}
\hrule